\documentclass[a4paper,11pt]{article}
\pdfoutput=1

\usepackage{jcappub}

\title{The Interplay between\\ the Dark Matter Axion\\ and Primordial Black Holes}

\author[a]{Kratika Mazde}
\author[b, c]{and Luca Visinelli}
\affiliation[a]{School of Physics, Indian Institute of Science Education and Research,\\ Maruthamala PO, Vithura, Thiruvananthapuram, 695551, Kerala, India}
\affiliation[b]{Astronomy division, Tsung-Dao Lee Institute (TDLI),\\ 520 Shengrong Road, 201210 Shanghai, P.\ R.\ China}
\affiliation[c]{School of Physics and Astronomy, Shanghai Jiao Tong University,\\ 800 Dongchuan Road, 200240 Shanghai, P.\ R.\ China}
\emailAdd{kratikamazde18@iisertvm.ac.in}
\emailAdd{luca.visinelli@sjtu.edu.cn}

\abstract{If primordial black holes (PBHs) had come to dominate the energy density of the early Universe when oscillations in the axion field began, we show that the relic abundance and expected mass range of the QCD axion would be greatly modified. Since the QCD axion is a potential candidate for dark matter (DM), we refer to it as the DM axion. We predominantly explore PBHs in the mass range $(10^6 - 5\times 10^8)\,$g. We investigate the relation between the relic abundance of DM axions and the parameter space of PBHs. We numerically solve the set of Boltzmann equations, that governs the cosmological evolution during both radiation and PBH-dominated epochs, providing the bulk energy content of the early Universe. We further solve the equation of motion of the DM axion field to obtain its present abundance. Alongside non-relativistic production mechanisms, light QCD axions are generated from evaporating PBHs through the Hawking mechanism and could make up a fraction of the dark radiation (DR). If the QCD axion is ever discovered, it will give us insight into the early Universe and probe into the physics of the PBH-dominated era. We estimate the bounds on the model from DR axions produced via PBH evaporation and thermal decoupling, and we account for isocurvature bounds for the period of inflation where the Peccei-Quinn symmetry is broken. We assess the results obtained against the available CMB data and we comment on the forecasts from gravitational wave searches. We briefly state the consequences of PBH accretion and the uncertainties this may further add to cosmology and astroparticle physics modeling.}

\begin{document}
\maketitle
\flushbottom

\section{Introduction}
\label{sec:introduction}

Black Holes (BHs) are an elegant solution to Einstein's field equations and an extreme consequence of the general theory of relativity. A vast array of astrophysical evidence for the existence of BHs in the Universe has mounted over the last couple of decades, which includes the motion of S stars around the compact region of Sagittarius A at the center of our Galaxy~\cite{Ghez:1998ph, Ghez:2008ms, Gillessen:2008qv}, the detection of gravitational waves (GWs) from binary BH coalescence~\cite{LIGOScientific:2016aoc, LIGOScientific:2018mvr}, and the imaging of the dark shadow surrounded by the luminous ring forming around supermassive BHs residing within galactic centers~\cite{EventHorizonTelescope:2019dse, EventHorizonTelescope:2022xnr}. BHs form in astrophysical environments when a substantially massive body collapses within a small region of space. This occurs at the end of the life cycles of massive stars, see Ref.~\cite{Heger:2002by} for an in-depth understanding.

The evidence mentioned above supports the existence of astrophysical BHs. In principle, BHs could have also formed primordially through gravitational collapses, through large density fluctuations in the plasma. This essentially takes place whenever the amplitude of a mode, while re-entering the horizon, crosses the threshold value for gravitational collapse well before the reionization epoch~\cite{Carr:1974nx, Carr:1975qj, Zeldovich:1967lct}. From a Press-Schechter approach, the PBH mass fraction at formation is obtained by integrating the probability density function for density perturbations above the critical value $\delta_c \approx 0.4\textrm{--}0.7$~\cite{Shibata:1999zs, Niemeyer:1999ak, Musco:2004ak}.\footnote{An alternative approach based on peak statistics leads to similar results under certain assumptions~\cite{Wu:2020ilx}. A detailed comparison among various approaches is demonstrated in Ref.~\cite{Gow:2020bzo}.} These PBHs might have also formed through curvature perturbations~\cite{Garcia-Bellido:1996mdl}, from long-range forces mediated by the scalar fields~\cite{Flores:2020drq}, from bubbles~\cite{Maeso:2021xvl}, or from a change in the equation of state of the primordial plasma due - in large part - to a phase transition~\cite{Byrnes:2018clq}. Detailed reviews on the subject are given in Refs.~\cite{Carr:2020gox, Carr:2020xqk}. Studying how these PBHs affect the present-day Universe and finding a signature for their existence is one of the primary goals of research in cosmology. 

Owing to Hawking radiation, BHs irradiate as ideal black bodies across a spectrum peaking at the Hawking temperature ($T_H$)~\cite{Hawking:1974rv}, effectively losing a portion of their mass in the process. For instance, PBHs with masses $M_{\rm BH} \gtrsim 5\times 10^{14}\,$g have a lifetime that exceeds the age of the Universe and could, in principle, account for the dark matter (DM) in the Universe. Several PBH mass windows have been scrutinized by various probes such as assessments against temperature and polarization data of the cosmic microwave background (CMB)~\cite{Ali-Haimoud:2016mbv}, the CMB anisotropy power spectrum~\cite{Acharya:2020jbv}, microlensing~\cite{Niikura:2017zjd}, and dynamical constraints~\cite{Carr:1997cn, Carr:2020erq}. Asteroids of masses $M \sim 10^{16}\,$g can be explored by studying PBHs residing in nearby structures and evaporating into soft gamma rays~\cite{Coogan:2020tuf}. See Refs.~\cite{Carr:2016drx, Green:2020jor} for reviews.

Light PBHs with masses $M_{\rm BH} \lesssim 5\times 10^8\,$g had a negligible impact on the CMB spectrum or BBN since their production and evaporation occurred pre-BBN.\footnote{Scenarios for post-BBN production have also been recently explored~\cite{Chakraborty:2022mwu}.} PBHs evaporate and emit information as radiation through Hawking mechanism and this phenomenon could alleviate various issues persisting in cosmology, like the occurrence of baryon asymmetry~\cite{Toussaint:1978br, Turner:1979zj} and tension in the Hubble constant~\cite{Nesseris:2019fwr, DiValentino:2021izs}. These light PBHs might have also altered the history of the Universe prior to BBN and influenced DM production. Furthermore, the observed DM in the Universe is composed of yet another component: an exotic new light field. Weakly interacting massive particles (WIMPs) could also be potential DM candidates~\cite{Jungman:1995df}. The evaporation of PBHs might have been the main driver for the production of the said potential DM candidate~\cite{Fujita:2014hha, Allahverdi:2017sks, Lennon:2017tqq, Morrison:2018xla, Hooper:2019gtx, Masina:2020xhk, Cheek:2021cfe, Chattopadhyay:2022fwa}, simultaneously impacting the DM production via misalignment mechanism in the early Universe~\cite{Baldes:2020nuv, Gondolo:2020uqv}. The interplay between PBHs and DM halos around them can lead to a multitude of constraints on BH and DM formation models~\cite{Boucenna:2017ghj, Bertone:2019vsk, Hertzberg:2020hsz, Carr:2020mqm, Gines:2022qzy, Chanda:2022hls}.

Here, we consider a scenario where PBHs incidentally affect the formation of DM by temporarily meddling with the thermal history of the Universe. As a matter of fact, the phenomenon of PBHs evaporation could explain a period of early matter domination prior to BBN, during which the properties of DM relics such as their abundance and free-streaming velocity would have been substantially modified~\cite{Chung:1998rq, Moroi:1999zb, Gelmini:2008sh, Visinelli:2015eka, Visinelli:2017qga}. We study the QCD axion~\cite{Weinberg:1977ma, Wilczek:1977pj}, a potential DM candidate and a hypothetical light particle whose existence is postulated as the solution to the strong-CP problem. It was proposed by Peccei and Quinn (PQ)~\cite{Peccei:1977ur, Peccei:1977hh}. Once the Hubble expansion rate falls below the axion mass, the coherent oscillations stored in the axion field behave as matter and can account for a fraction or even the totality of the DM observed~\cite{Preskill:1982cy, Abbott:1982af, Dine:1982ah}. The period of early matter domination (resulting from the presence of PBHs) would have an influence on the onset of field oscillations and can lead to a different DM abundance at present for a given axion mass~\cite{Lazarides:1987zf, Lazarides:1990xp, Visinelli:2009kt, Visinelli:2017imh}, which, in turn, opens up the possibility for a lighter DM axion than what is expected in the standard cosmological scenario, with observable consequences in theoretical astrophysics and in the laboratory~\cite{Visinelli:2018wza, Nelson:2018via, Ramberg:2019dgi, Ramberg:2020oct, Arias:2021rer, Bao:2022hsg}.\footnote{Alternative cosmologies other than early matter domination have also been suggested~\cite{Arias:2022qjt}.} This circumstance is of interest to us in the PBH mass window $M_{\rm BH} = (10^6 - 5\times 10^8)\,$g. We obtain a more elaborate picture of this as we proceed.

We advance over previous estimates of axion abundance in the presence of PBHs~\cite{Bernal:2021yyb} by implementing a series of novel approaches. The equation of motion for the axion is solved numerically using the plasma temperature as the independent variable, to facilitate the inclusion of temperature-dependent quantities. We also present a detailed numerical analysis of the Boltzmann equations which govern the abundance of the energy densities in the radiation bath and in the PBH-dominated background, similarly to Ref.~\cite{Bernal:2021bbv}. Our results show that if PBHs did dominate the energy density of the early Universe around the period of QCD phase transition when the axion field oscillations began, the range of mass allowed for a DM axion is widened with respect to what is allowed in the standard cosmology. 

We explore the bounds from thermal axions, generated by PBH evaporation as dark radiation, and we account for axion isocurvature bounds in the case where the PQ symmetry is broken during the period of inflation. We study and comment on the changes our current cosmological model might undergo should PBH accretion be taken into consideration. We assess the bounds on the model parameters by testing them against the CMB and GW data secured from experiments conducted so far. Our novel observations may serve to probe into the physics of a plethora of cosmological models that can be mapped onto our minimal setup.

The paper is organized as follows. In section~\ref{sec:PBHevap} we provide an overview of PBH evaporation. In section~\ref{sec:backgroundeqs} we discuss the cosmological setup which we later study and solve equations over numerically. In section~\ref{sec:axionreview} we review the theory of the axion. In section~\ref{sec:method} we present the methods adopted to solve the set of equations numerically and we obtain certain results, which are stated in section~\ref{sec:results}. The results are discussed in section~\ref{sec:discussion} and conclusions are given in section~\ref{sec:conclusions}. We work with natural units, i.e., $\hbar = c = k_B = 1$, and we refer to the Planck mass as $m_{\rm Pl} = G_N^{-1/2}$. The open source code is available at \href{https://github.com/lucavisinelli/AXIONPBH}{github.com/lucavisinelli/AXIONPBH}.

\section{Primordial black holes}
\label{sec:PBHevap}

\subsection{Formation and abundance}
\label{sec:formation} 

The abundance of PBHs is defined in terms of the fraction of matter-energy that undergoes gravitational collapse at the time of formation $t_f$,
\begin{equation}
    \label{eq:definebeta}
    \beta = \frac{\rho_{\rm BH}(t_f)}{\rho_{\rm crit}(t_f)}\,,
\end{equation}
where the subscript ``$f$'' stands for the formation time, $\rho_{\rm BH}$ is the PBH energy density, and $\rho_{\rm crit}(t) \equiv 3H^2(t)/(8\pi G_N)$ is the critical energy density at time $t$ in terms of the Hubble rate $H(t)$ and Newton's constant $G_N$. Since we assume the formation occurs during the radiation domination and with a monochromatic PBH mass spectrum, the ratio of the initial PBH mass to the horizon mass is given by $\gamma \approx 0.2$~\cite{Carr:1975qj}, so that the PBH mass is
\begin{equation}
    \label{eq:initialmass}
    M_{\rm BH} = \gamma \frac{4\pi}{3}\frac{\rho_{{\rm crit}, f}}{H_f^3} = \frac{\gamma}{2G_N H_f}\,,
\end{equation}
where $H_f \equiv H(t_f)$ and the critical density at formation is $\rho_{\rm crit}(t_f) = 3\gamma^2/(32 \pi G_N^3 M_{\rm BH}^2)$. We assume that the Universe is radiation-dominated at $t_f$, with energy density in the plasma
\begin{equation}
    \label{eq:rhorad}
    \rho_r(T) = \frac{\pi^2}{30}g_*(T)T^4\,,
\end{equation}
and the function $g_*(T)$ gives the number of relativistic degrees of freedom at temperature $T$. Under these assumptions, the initial temperature of the primordial plasma surrounding the PBH is
\begin{equation}
    \label{eq:Tinitial}
    T_f = \left(\frac{45}{16 \pi^3 G_N^3 M_{\rm BH}^2} \frac{\gamma^2}{g_*(T_f)}\right)^{1/4}\,.
\end{equation}
In what follows, we assume that PBHs do not carry a charge and have negligible spin~\cite{Mirbabayi:2019uph, DeLuca:2019buf}. We consider a monochromatic PBH mass function, described by the expression in Eq.~\eqref{eq:initialmass} which relates the mass of the PBH produced to cosmic time, and which has been explored extensively in previous literature, see e.g.\ Ref.~\cite{Carr:2016drx} for additional details.

Being formed in an early radiation-dominated epoch, PBHs would dilute their energy density much slower than the surrounding radiation, so that if $\beta$ is sufficiently large (see Eq.~\eqref{eq:lowerbound1} below) they would come to dominate the expansion rate until their eventual evaporation. The initial abundance of PBHs is tracked by the BH yield $Y_{\rm BH}(T) \equiv n_{\rm BH}(T)/s(T)$, where $n_{\rm BH} = \rho_{\rm BH}/M_{\rm BH}$ and the entropy density $s(T) \equiv (2\pi^2/45)g_S(T)T^3$ is defined in terms of the entropy degrees of freedom $g_S(T)$ at temperature $T$. The PBH yield at formation is
\begin{equation}
    Y_{\rm BH}(t_f) = \frac{3 T_f}{4M_{\rm BH}} \beta\,,
\end{equation}
and with this, the energy density in PBHs grows to match the energy density of relativistic particles at the time of PBH-radiation equality $t_{\rm EQ}$. In the absence of an entropy dilution during an early radiation dominated epoch, the PBH energy density at $t_{\rm EQ}$ is $\rho_{\rm BH} = Y_{\rm BH}(t_f)\,M_{\rm BH}\,s(T_{\rm EQ})$, which gives us the temperature value for PBH-radiation equality. Given below is the said temperature value.
\begin{equation}
    \label{eq:equality}
    T_{\rm EQ} \approx \beta T_f\,.
\end{equation}

\subsection{Physical scale}

PBHs could have resulted from the amplification of overdensities during inflation at a specific scale $k$~\cite{Garcia-Bellido:1996mdl}. We discuss the relation between the PBH mass at formation $M_{\rm BH}$ and physical scale $k$ (at formation), in order to gain a better understanding of the new physics required during inflation to produce the desired features. If the post-reheating Universe is radiation-dominated, the mass enclosed within a horizon of size $k^{-1}$ is
\begin{equation}
    M(k) = M_{\rm rh}\,(k/k_{\rm rh})^{-2}\,,
\end{equation}
where the horizon mass at reheating time depends on the reheating temperature $T_{\rm rh}$ as ($m_{\rm Pl} = G_N^{-1/2}$)
\begin{equation}
    M_{\rm rh} = \frac{4\pi}{3} \frac{\rho_{\rm rh}}{H_{\rm rh}^3} = \left(\frac{45}{16\pi^2g_*(T_{\rm rh})}\right)^{1/2}\frac{m_{\rm Pl}^3}{T_{\rm rh}^2} \approx 2\times 10^6{\rm\,g}\left(\frac{T_{\rm rh}}{10^{13}{\rm\,GeV}}\right)^{-2}\left(\frac{g_*(T_{\rm rh})}{106.75}\right)^{-1/2}\,,
\end{equation}
while the corresponding physical wave number depends on the {\it Planck} pivot scale $k_{0.05} = 0.05\,{\rm Mpc}^{-1}$ as
\begin{equation}
    \label{eq:krh0}
    k_{\rm rh} = k_{0.05}\,e^{\Delta N_{0.05}}\,\left(\frac{g_*(T_{\rm rh})}{g_*(T_{0.05})}\right)^{-1/6}\,.
\end{equation}
Here, $\Delta N_{0.05}$ is the number of $e$-folds that take place from the onset of the radiation phase until the reentry of the scale $k_{0.05}$, when the relativistic degrees of freedom are $g_*(T_{0.05})$, and it is related to the details of inflation by~\cite{Dalianis:2018ymb}
\begin{equation}
    N_e + \ln\left(\frac{H_I}{H_{0.05}}\right) = \frac{1}{2}N_{\rm rh} + \Delta N_{0.05}\,,
\end{equation}
where $H_I$ is the scale of inflation, $N_{\rm rh}$ is the number of $e$-folds during the reheating stage, and~\cite{Liddle:2003as}
\begin{eqnarray}
    N_e &=& \left[\ln\left(\frac{a_{\rm eq}H_{\rm eq}}{a_0 H_0}\right) + \frac{1}{4}\ln\left(\frac{m_{\rm Pl}^4}{\rho_{\rm eq}}\right)\right] + \frac{1}{4}\ln\left(\frac{V_{0.05}^2}{\rho_I\,m_{\rm Pl}^4}\right) + \frac{1}{12}\ln\frac{\rho_{\rm rh}}{\rho_I}\,,
\end{eqnarray}
where, setting the central value results from the latest {\it Planck} measurements~\cite{Planck:2018jri} for $\Omega_m \approx 0.3153$, $\Omega_m h^2 \approx 0.1430$, and the redshift at matter-radiation equality $z_{\rm eq} \approx 3402$, the first two terms in the square brackets yield the value
\begin{equation}
    \ln\left(\frac{a_{\rm eq}H_{\rm eq}}{a_0 H_0}\right) + \frac{1}{4}\ln\left(\frac{m_{\rm Pl}^4}{\rho_{\rm eq}}\right) = \frac{1}{2}\ln\left(\Omega_m (1 + z_{\rm eq})\right) + \frac{1}{4}\ln\left(\frac{8\pi m_{\rm Pl}^2}{3 \Omega_m H_0^2 (1 + z_{\rm eq})^3}\right) \approx 68.3\,.   
\end{equation}
Combining these last two expressions gives
\begin{equation}
    \label{eq:deltaN005}
    \Delta N_{0.05} = 68.3 + \frac{1}{4}\ln\left(\frac{V_{0.05}^2}{\rho_I\,m_{\rm Pl}^4}\right) + \ln\left(\frac{H_I}{H_{0.05}}\right) - \frac{3}{4}N_{\rm rh}\,,
\end{equation}
where we used the relation $\rho_{\rm rh} = \rho_I e^{-3N_{\rm rh}}$; valid for the matter-dominated stage during reheating and can be recast in terms of $T_{\rm rh}$ as
\begin{equation}
    \label{eq:Nrh}
    N_{\rm rh} = -\frac{4}{3}\ln\left[\left(\frac{4\pi^3}{45}g_*(T_{\rm rh})\right)^{1/4}\frac{T_{\rm rh}}{(m_{\rm Pl}\,H_I)^{1/2}}\right]\,.
\end{equation}
Finally, combining Eqs.~\eqref{eq:krh0},~\eqref{eq:deltaN005} and~\eqref{eq:Nrh} gives
\begin{eqnarray}
    \label{eq:krh}
    k_{\rm rh} &=& k_{0.05}\,\frac{V_{0.05}^{1/2}}{\rho_I^{1/4}\,m_{\rm Pl}} \frac{H_I}{H_{0.05}}\left(\frac{4\pi^3}{45}g_*(T_{\rm rh})\right)^{1/4}\frac{T_{\rm rh}}{(m_{\rm Pl}\,H_I)^{1/2}}\,\left(\frac{g_*(T_{\rm rh})}{g_*(T_{0.05})}\right)^{-1/6}\,e^{68.3}\\
    &=& k_{0.05}\left(\frac{\pi^2}{30}g_*(T_{\rm rh})\right)^{1/4}\frac{T_{\rm rh}}{m_{\rm Pl}}\,\left(\frac{g_*(T_{\rm rh})}{g_*(T_{0.05})}\right)^{-1/6}\,e^{68.3} \approx 4\times 10^{22}{\rm\,Mpc^{-1}}\left(\frac{T_{\rm rh}}{10^{13}{\rm\,GeV}}\right)\,,\nonumber
\end{eqnarray}
The physical scale at which the PBH of mass $M_{\rm BH}$ forms is then (see also Ref.~\cite{Ballesteros:2017fsr})
\begin{equation}
    k_{\rm BH} \approx 10^{22}{\rm\,Mpc}^{-1}\,\left(\frac{M_{\rm BH}}{10^7{\rm\,g}}\right)^{-1/2}\,,
\end{equation}
where we used Eq.~\eqref{eq:initialmass} with $\gamma = 0.2$. On scales such as this, the primordial power spectrum is very loosely constrained (see e.g.\ Ref.~\cite{Clesse:2015wea}) so that PBHs can be effectively produced and dominate the expansion rate for some time.

Possible mechanisms of production include the inflaton potential with a flat-inflection feature or a multi-phase inflation model. More in detail, the first example consists of the ultra-slow roll inflation model~\cite{Kinney:2005vj, Motohashi:2017kbs}, which leads to an amplification of the fluctuations on specific scales. Explicit realizations for a PBH mass spectrum within the ultra-slow roll phase have been discussed in Refs.~\cite{Dalianis:2018frf, Wu:2021mwy}.

\subsection{Evaporation}
\label{sec:evaporation}

According to the Hawking mechanism, a BH of mass $M_{\rm BH}$ radiates a spectrum of particles lighter than the temperature~\cite{Hawking:1975vcx}.
\begin{equation}
    \label{eq:Hawking}
	T_H = \frac{1}{8\pi G_N M_{\rm BH}}\,,
\end{equation}
The evolution of a non-rotating BH\footnote{Several numerical tools are further available to compute the evaporation rate in different scenarios. A detailed list is given in Ref.~\cite{Auffinger:2022khh}.} is determined by its mass loss due to the emitted particles~\cite{Hawking:1975vcx, Page:1976wx, Carr:1976zz}, which accounts for the sum over all emitted species $i$ with spin $s_i$ and orbital angular momentum $\ell$.
\begin{equation}
    \frac{{\rm d}M_{\rm BH}}{{\rm d}t} = -\sum_{i,\ell} \frac{g_i}{2\pi} \int_0^{+\infty}\frac{\omega\, \Gamma_{s_i,\ell}(\omega)}{e^{\omega/T_H} \pm 1}{\rm d}\omega\,,
\end{equation}
where $g_i$ denotes the number of degrees of freedom for the particle $i$~\cite{MacGibbon:1990zk, MacGibbon:1991tj} and the factor $\Gamma_{s_i,\ell}(\omega)$ is the probability of a particle produced near the horizon with an energy $\omega$ and quantum numbers $s_i$, $\ell$ escaping to infinity, or conversely the probability of absorbing an incoming wave of the same energy. The absorption probability $\Gamma_{s_i,\ell}(\omega)$ is obtained by decomposing the particle wave function into spherical harmonics, with the resulting equation being the Teukolsky equation~\cite{Teukolsky:1973ha} for the radial wave function $R(r)$. When applied to a non-rotating BH, the Teukolsky equation reduces to a Schr{\"o}dinger-like expression known as the Bardeen-Press equation~\cite{Bardeen:1973xb}. The numerical computation has been carried out in Ref.~\cite{Page:1976wx} which leads us to the below expression for the mass loss rate 
\begin{equation}
	\label{eq:masslossrate}
	\frac{{\rm d}M_{\rm BH}/{\rm d}t}{M_{\rm BH}}\bigg|_{\rm evap} = -\frac{\mathcal{G}\,g_H(T_H)}{30720\pi G_N^2M_{\rm BH}^3}\,,
\end{equation}
where $\mathcal{G} \approx 3.8$ is the BH graybody factor. The spin-weighted number of degrees of freedom $g_H(T_H)$ accounts for all existing particles with mass less than the Hubble temperature ($T_H$).
\begin{equation}
    g_H(T_H) = \sum_i g_i \, w_{s_i}\,,
\end{equation}
where the coefficients $w_{s_i}$ are given as: $w_0 = 1.82$, $w_1 = 0.41$, $w_2 = 0.05$, and $w_{1/2} = 1$ (neutral) or $w_{1/2} = 0.97$ (absolute elementary charge). Note, the effect of mutual repulsion between charged particles is accounted for even though the BH is not charged~\cite{Page:1977um}. Summing all contributions from SM particles gives $g_H^{\rm SM} \simeq 102.6$.  With the inclusion of an additional scalar degree of freedom, we obtain $g_H \simeq 104.5$. We refer to this latter value of $g_H$ for SM+axion content, neglecting any and all changes caused by temperature. Because the Hawking temperature $T_H$ for BHs that evaporate before BBN is well above the electroweak scale, all SM particles are emitted at this stage.

For cases where $g_H$ does not change with time, we solve Eq.~\eqref{eq:masslossrate} and obtain the characteristic time within which a PBH evaporates:
\begin{equation}
    \label{eq:tevap}
    t_{\rm evap} = 30760\pi\, G_N^2M^3_{\rm BH}/(\mathcal{G}\,g_H)\,.
\end{equation}
Here, we neglect the temporal change in the PBH mass since it only affects the endmost stages of the PBH evolution. Demanding that the corresponding temperature at PBH evaporation be $T_{\rm evap} > T_{\rm BBN}$ and fixing $T_{\rm BBN} \sim 1\,$MeV~\cite{Kawasaki:1999na, Kawasaki:2000en, Hannestad:2004px, Ichikawa:2005vw} leads to the bound
\begin{equation}
    \label{eq:BBNconstraint}
    M_{\rm BH} \lesssim 5\times 10^{8}{\rm \,g}\,.
\end{equation}
Consistency in the successful results obtained accounting BBN in a radiation-dominated background demands that any potential early population of PBHs should take over radiation prior to BBN. Further demanding that a period of PBH domination effectively occurs at some point and a sizeable population of PBHs with a sufficiently large value of $\beta$ is produced, $T_{\rm EQ} > T_{\rm evap}$ with $T_{\rm EQ}$ in Eq.~\eqref{eq:equality}, gives
\begin{equation}
    \label{eq:lowerbound1}
    \beta \gtrsim \left(\sqrt{\frac{g_*(T_f)}{g_*(T_{\rm evap})}}\,\frac{G_N M_{\rm BH}}{\gamma\, t_{\rm evap}}\right)^{1/2}\,,
\end{equation}
or $\beta \gtrsim 5\times 10^{-10}(10^4{\rm\,g}/M_{\rm BH})$. A different lower bound is obtained by suggesting that PBH domination effectively terminates before the initiation of BBN. This is equivalent to demanding that $T_{\rm EQ} > T_{\rm BBN}$, or
\begin{equation}
    \label{eq:lowerbound2}
    \beta \gtrsim 5\times 10^{-17} \,\left(\frac{M_{\rm BH}}{10^4{\rm\, g}}\right)^{1/2}\,.
\end{equation}
Alongside the lower bounds on the fractional PBH abundance, an upper bound on $\beta$ is also obtained. We find the upper bound demanding that GWs produced from PBH isocurvature modes do not spoil BBN~\cite{Domenech:2020ssp}, as we revise later in section~\ref{sec:PBHisobounds}.

\section{Evolution of PBHs in cosmology}
\label{sec:backgroundeqs}

\subsection{Boltzmann equations}

Since PBHs are formed with a negligible momentum in the comoving coordinate frame, they behave as matter with Boltzmann equation for energy density as
\begin{equation}
    \label{eq:PBHmassloss}
    \frac{{\rm d}\rho_{\rm BH}}{{\rm d}t} + 3H\rho_{\rm BH} = \frac{{\rm d}M_{\rm BH}/{\rm d}t}{M_{\rm BH}}\, \rho_{\rm BH}\,,
\end{equation}
where the BH mass loss rate due to evaporation is given in Eq.~\eqref{eq:masslossrate} and the Hubble rate is found from the following Friedmann equation 
\begin{equation}
    \label{eq:hubble}
    H^2 = \frac{8\pi G_N}{3}\left(\rho_r +\rho_{\rm BH}\right)\,.
\end{equation}
The by-products of the BH evaporation contribute to the radiation energy density as
\begin{equation}
    \label{eq:rad}
    \frac{{\rm d}\rho_r}{{\rm d}t} + 4H\rho_r = -\frac{{\rm d}M_{\rm BH}/{\rm d}t}{M_{\rm BH}} \,\rho_{\rm BH}\,,
\end{equation}
where the radiation energy density tracks the plasma temperature as given in Eq.~\eqref{eq:rhorad}. In Appendix~\ref{sec:dof}, we outline the scheme we deploy to compute the function $g_*(T)$ and we provide a numerical fit for the SM case.

A consequence of Eq.~\eqref{eq:rad} is seen in the non-conservation front of the entropy during the evaporation phase, which can be stated in terms of the evolution of the entropy density $s(T)$ as
\begin{equation}
    \label{eq:entropy}
    \frac{{\rm d} s}{{\rm d} t} + 3 H s = -\frac{{\rm d}M_{\rm BH}/{\rm d}t}{M_{\rm BH}}\,\frac{\rho_{\rm BH}}{T}\,.
\end{equation}
Since Eq.~\eqref{eq:entropy} is applied at temperatures above $T_{\rm BBN}$, we have neglected the difference between $g_S(T)$ and $g_*(T)$. It comes in handy to express Eq.~\eqref{eq:PBHmassloss} with the temperature being the independent variable instead of time, in combination with Eq.~\eqref{eq:rad} and the definition of $\rho_r$, to obtain
\begin{equation}
    \label{eq:rhovsT}
    Q(T)\frac{{\rm d} \rho_{\rm BH}}{{\rm d}T} + \frac{3}{T}\frac{H}{H_{\rm eff}}\rho_{\rm BH} = -\frac{{\rm d}M_{\rm BH}/{\rm d}t}{M_{\rm BH}}\frac{\rho_{\rm BH}}{T\,H_{\rm eff}}\,,
\end{equation}
where the effective Hubble rate is
\begin{equation}
    H_{\rm eff} = H + \frac{{\rm d}M_{\rm BH}/{\rm d}t}{M_{\rm BH}}\frac{\rho_{\rm BH}}{4\rho_r}\,.
\end{equation}
Eq.~\eqref{eq:rhovsT} coincides with Eq.~(35) in Ref.~\cite{Masina:2020xhk} except for the presence of the function $Q(T)$ accounting for the change in the effective number of relativistic degrees of freedom as
\begin{equation}
    \label{def:QT}
    Q(T) \equiv \left(1 + \frac{T}{4g_*(T)}\frac{{\rm d} g_*(T)}{{\rm d}T}\right)^{-1}\,.
\end{equation}
We discuss in detail the modified Eq.~\eqref{eq:rhovsT} we use to derive the results for the background in Appendix~\ref{sec:numerics}.

\begin{figure}
    \centering
    \includegraphics[width=0.8\linewidth]{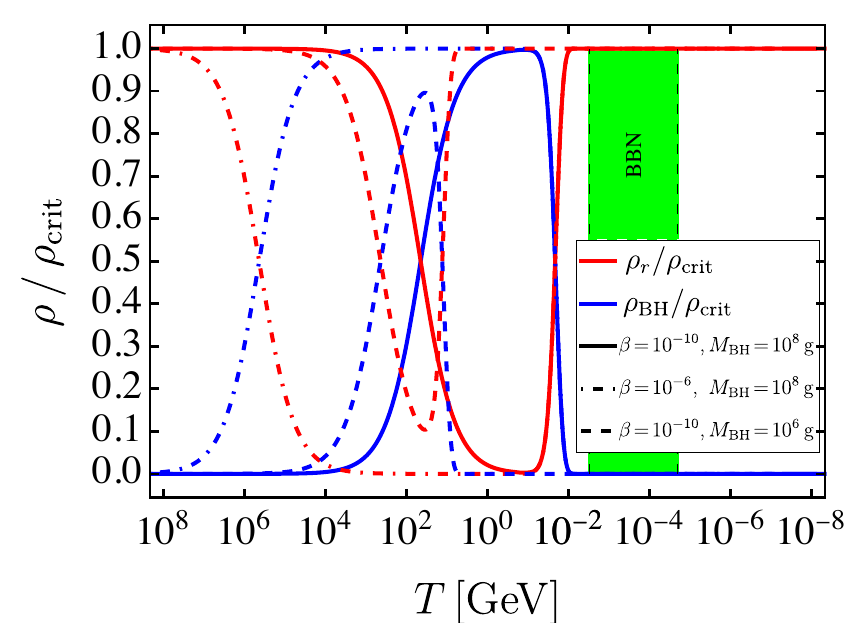}
    \caption{The fractional abundances $\rho_{\rm BH}/\rho_{\rm crit}$ (blue) and $\rho_r/\rho_{\rm crit}$ (red), where $\rho_{\rm crit}$ is the critical density at any given time, as a function of the plasma temperature $T$ in GeV. The initial fractional abundance and mass of the PBH component is $(\beta, M_{\rm BH}) = (10^{-10}, 10^8{\rm\,g})$ (solid line), $(\beta, M_{\rm BH}) = (10^{-6}, 10^8{\rm\,g})$ (dot-dashed line), and $(\beta, M_{\rm BH}) = (10^{-10}, 10^6{\rm\,g})$ (dashed line), respectively.}
    \label{fig:BKGD}
\end{figure}

Note, the system of equations above is completely determined only after the set of parameters $(\beta, M_{\rm BH})$ is specified. For some values of the initial conditions $(\beta, M_{\rm BH})$, the PBH energy density temporarily dominates the expansion rate of the early Universe. For an illustrative purpose, in figure~\ref{fig:BKGD} we show the results obtained from solving Eq.~\eqref{eq:rhovsT} in terms of the quantities $\rho_{\rm BH}/\rho_{\rm crit}$ (blue) and $\rho_r/\rho_{\rm crit}$ (red), where $\rho_{\rm crit}$ is the critical density at any given time, as a function of the plasma temperature (GeV) derived from Eq.~\eqref{eq:rhorad}. Results are obtained for three distinct PBH masses $M_{\rm BH}$ with initial fractional abundances $\beta$, namely $(\beta, M_{\rm BH}) = (10^{-10}, 10^8{\rm\,g})$ (solid line), $(\beta, M_{\rm BH}) = (10^{-6}, 10^8{\rm\,g})$ (dot-dashed line), and $(\beta, M_{\rm BH}) = (10^{-10}, 10^6{\rm\,g})$ (dashed line) respectively. The PBH mass $M_{\rm BH}$ regulates the phenomenon of PBH evaporation, while the fractional abundance $\beta$ determines when the transition from early radiation to early matter domination begins, i.e., if it ever does. Moreover, we ensure that the temperature of the Universe at BBN initiation is $T_{\rm BBN} \sim 1\,$MeV. By the time BBN starts, and radiation is back to ruling the energy density of the Universe, PBHs have fully decayed and the abundance of light elements has been fixed. BBN lasts approximately 180 seconds~\cite{Workman:2022ynf}. 

\subsection{Assessing the distortion of the mass spectrum}
\label{sec:accretion}

In this work, we present results for the case in which the distribution of PBH masses is monochromatic at formation. Even in this illustrative scenario, it could be possible that post-formation effects distort such a distribution by shifting some of the PBH masses away from the initial value. For this, we discuss the role of binary mergers and accretion, showing that these two effects can be neglected for the range of masses considered. In section~\ref{sec:extendedmassdistribution} we also discuss the role of an extended distribution of masses.

We first discuss PBH merging rate, which proceeds hierarchically starting from a spatially uncorrelated distribution of PBHs initially formed at rest: closer PBHs would merge first, followed by those that lie further apart. The release of GWs might help PBH pairs in otherwise hyperbolic trajectories to lose enough energy so that they end up in bound orbits. Given the initial abundance $\beta$, PBH merging becomes important at their mutual separation $x$ for which their energy density within a volume of radius $x$ is greater than the average density of the Universe~\cite{Nakamura:1997sm}. Given the mean comoving distance at PBH-radiation equality as $\bar x_{\rm EQ} \equiv (2M_{\rm BH}/\rho_{\rm EQ})^{1/3}$, the PBH  energy density is then
\begin{equation}
    \rho_{\rm BH} = \rho_{\rm EQ}\left(\frac{\bar x_{\rm EQ}}{x a}\right)^3\,,
\end{equation}
while the radiation energy density scales as $\rho_r = \rho_{\rm BH}(a_{\rm EQ}/a)^4$, so that two PBHs at a physical separation $x$ begin to fall onto each other at the scale factor $a_m \equiv (a_{\rm EQ})^4 (x/\bar x_{\rm EQ})^3$. The coalescence time for binary PBHs in circular orbits and at a distance $\bar x_{\rm EQ}$ at $t_{\rm EQ}$ is~\cite{Zagorac:2019ekv}
\begin{equation}
    t_c \simeq \sqrt{\frac{32 \pi}{3}}\frac{5}{16}\left(\frac{\pi}{6}\right)^{5/6} \beta^{-16/3} G_N M_{\rm BH}\,, 
\end{equation}
which correctly increase with decreasing $\beta$, so that below some critical value $\bar\beta(M_{\rm BH})$, the coalescence time becomes greater than the evaporation time in Eq.~\eqref{eq:tevap}. For $M_{\rm BH} = 10^6\,$g ($M_{\rm BH} = 10^8\,$g), this critical value is equal to $\bar\beta = 4\times 10^{-5}$ ($\bar\beta = 6\times 10^{-6}$), practically covering the whole of the parameter space of interest for the results in this work, see figure~\ref{fig:densityplot} below. We conclude that, for the light PBHs evaporating prior to BBN, merging does not appreciably modify the mass distribution.

We now turn to accretion. PBHs accrete mass from the surrounding plasma at a rate~\cite{Bondi:1952ni}
\begin{equation}
    \label{eq:accretion}
    \frac{{\rm d}M_{\rm BH}/{\rm d}t}{M_{\rm BH}}\bigg|_{\rm accr} = f_{\rm accr}\,4\pi r_{\rm BH}^2\frac{\rho_r}{M_{\rm BH}}\,,
\end{equation}
where $r_{\rm BH} = 2G_N M_{\rm BH}$ is the BH Schwarzschild radius and the accretion efficiency is $f_{\rm accr} = 27/16$ for the non-interacting plasma~\cite{1966AZh....43..758Z}. When accretion is included, the PBH mass loss term in Eqs.~\eqref{eq:PBHmassloss}--\eqref{eq:rad} modifies as
\begin{equation}
    \frac{{\rm d}M_{\rm BH}/{\rm d}t}{M_{\rm BH}} = \frac{{\rm d}M_{\rm BH}/{\rm d}t}{M_{\rm BH}}\bigg|_{\rm evap} + \frac{{\rm d}M_{\rm BH}/{\rm d}t}{M_{\rm BH}}\bigg|_{\rm accr}\,.
\end{equation}
Clearly, the two terms have opposite signs as accretion helps bring mass into the PBH while evaporation leads to a net mass ejection in the form of Hawking radiation. To better understand the role of accretion, we compare the two timescales at the time of PBH-radiation equality $T_{\rm EQ} = \beta T_f$, where the rates between accretion and evaporation read
\begin{equation}
    \label{eq:accretion1}
    \frac{t_{\rm evap}}{t_{\rm accr}}\bigg|_{T_{\rm EQ}} \approx \frac{46140\pi\,\gamma^2}{\mathcal{G}\,g_H}f_{\rm accr} \beta^4G_N M^2_{\rm BH}\frac{g_*(\beta T_f)}{g_*(T_f)}\,.
\end{equation}
In the mass ranges of PBHs that are relevant for this work, this ratio is always smaller than one and accretion can be safely neglected from the time when PBHs start dominating the energy density.

\subsection{Bounds from the effective number of neutrinos}
\label{sec:PBHisobounds}

Any excess in radiation at BBN or recombination can be translated into a contribution to the effective number of relativistic neutrinos $N_\nu$, following the definition for $g_*(T)$ predicting that the total number of relativistic degrees of freedom before neutrino decoupling at temperature $T_{\rm dec}$ be
\begin{equation}
    g_*(T \gtrsim T_{\rm dec}) = 2 + \frac{7}{8}(4 + 2N_\nu)\,,
\end{equation}
where $N_\nu = 2.99_{-0.33}^{+0.34}$ at 95\% confidence level (CL) using the TT, TE, EE + lowE combination of the angular correlations measured by the {\it Planck} Collaboration jointly with a lensing mapping and baryon acoustic oscillation data~\cite{Planck:2018vyg}. Setting $N_\nu = N_\nu^{\rm SM} + \Delta N_\nu$ with $N_\nu^{\rm SM} = 3.046$ being the number predicted within the SM~\cite{Mangano:2001iu}, see also Refs.~\cite{deSalas:2016ztq, Akita:2020szl, Bennett:2020zkv}. The contribution from relativistic species other than neutrinos is then
\begin{equation}
    \label{eq:rhoDR}
    \rho_{\rm DR} = \frac{\pi^2}{30}\frac{7}{4}\Delta N_\nu\,T^4\,,
\end{equation}
Here, the additional number of neutrino species $\Delta N_\nu$ parametrizes all contributions that are relativistic at recombination. These include primordial GWs, which might have been produced in the early Universe through various mechanisms, as well as any other exotic relativistic species such as axions.

Here, we review the contributions to $\Delta N_\nu$ from GWs released due to curvature perturbations. We leave the discussion of the contributions from dark radiation to section~\ref{sec:thermalaxions}. From the cosmological viewpoint, the energy density in GWs behaves identically to radiation and redshifts as $\rho_{\rm GW} \propto a^{-4}$. In a cosmology where PBHs first temporarily govern the energy density of the Universe and then evaporate, isocurvature modes associated with the perturbations in the energy density of PBHs convert into GWs ~\cite{Papanikolaou:2020qtd}. GWs are produced via varied mechanisms: gravitons are emitted by the Hawking evaporation of second-order GWs associated with the PBH formation scale ~\cite{Inomata:2020lmk}, the GWs from merging PBH in binary configurations, all occurring at frequencies generally higher than the window of detection that can be probed with terrestrial interferometry techniques such as the LIGO-VIRGO-KAGRA consortium. A comprehensive review is given in Ref.~\cite{Dolgov:2011cq}.

We demand that the energy density of GWs at BBN does not exceed the amount from an exotic relativistic component, or $\rho_{\rm GW, BBN} < \rho_{\rm DR}$~\cite{Caprini:2018mtu}. Using the results for $\rho_{\rm GW, BBN}$ in a PBH-dominated cosmology returns the bound~\cite{Ando:2018qdb, Inomata:2019ivs, Domenech:2020ssp, Passaglia:2021jla, Domenech:2021and}
\begin{equation}
    \label{eq:boundbeta}
    \beta \lesssim 1.1\times 10^{-6}\left(\frac{g_*(T_{\rm evap})}{106.75}\right)^{1/16}\left(\frac{M_{\rm BH}}{10^4{\rm\,g}}\right)^{-17/24}\,.
\end{equation}
Note, this bound has been derived using the bound $\Delta N_\nu \lesssim 0.2$ which is more stringent than the constraint placed by the {\it Planck} Collaboration at 95\% CL, although such a numerical change does not modify the considerations dramatically.

\section{Theory of the axion}
\label{sec:axionreview}

\subsection{Models for the invisible axion}

The QCD axion is a light pseudoscalar particle whose existence would address the strong-CP problem, a long-standing open question within the established SM. The energy density stored in the coherent axion oscillations could also explain the dark matter in the Universe, provided the mass of the axion $m_0$ falls within a specific mass range. The most general embedment of the axion model within the SM includes one PQ complex scalar field $\sigma$, of which the axion is the angle component, along with new fermions that are charged under the PQ symmetry. For example, in the Kim-Shifman-Vainshtein-Zakharov (KSVZ) model~\cite{Kim:1979if, Shifman:1979if}, the scalar field has the PQ charge $+2$ and the new fermions are new SU(3) triplets with the PQ charge $+1$ for left-handed chirality and $-1$ for right-handed chirality. Another possibility is the inclusion of a Higgs doublet~\cite{Zhitnitsky:1980tq, Dine:1981rt}. See Ref.~\cite{DiLuzio:2020wdo} for reviews on more differing configurations.

At energies below the PQ scale $f_{\rm PQ}$ and the electroweak scale, the PQ symmetry is broken and the PQ field is decomposed as
\begin{equation}
    \sigma = \frac{1}{\sqrt{2}}\left(\rho + v\right)\,e^{-i\phi/f_{\rm PQ}}\,,
\end{equation}
where $v$ is the vacuum of the PQ theory, $\rho$ is the radial field, and the axion field $\phi$ appears as the angle component of $\sigma$. Integrating out the motion of $\rho$ leaves us with the effective Lagrangian describing the evolution of the axion field as
\begin{equation}
    \label{eq:Lagrangian}
    \mathcal{L} \supset \frac{1}{2}\partial^\mu\phi\partial_\mu\phi - V(\phi) + \frac{\phi}{f_{\rm PQ}}\frac{\alpha_s}{8\pi}\tilde G_a^{\mu\nu}G^a_{\mu\nu} + \mathcal{L}_{\rm int} + \mathcal{L}_{\rm SM}\,,
\end{equation}
where $\alpha_s$ is the strong coupling, $\mathcal{L}_{\rm SM}$ is the SM Lagrangian, $V(\phi)$ is the axion self-interaction potential, $G_a^{\mu\nu}$ is the gluon field with color index $a$ and with dual field $\tilde G_a^{\mu\nu}$, and $\mathcal{L}_{\rm int}$ describes the interaction of the axion with the SM fermions and the gauge fields other than the gluon fields. As an axion-gluon term is expected to appear in the QCD axion theory, a rotation of the quark mass matrix leads to the axion mass~\cite{Weinberg:1977ma}
\begin{equation}
    \label{eq:axionmass}
    m_0 = \frac{\sqrt{z}}{1+z}\,\frac{m_\pi f_\pi}{f_{\rm PQ}} \equiv \frac{\Lambda^2}{f_{\rm PQ}}\,,
\end{equation}
where $z = 0.48$ is the ratio of masses of the up and down quarks, $m_\pi$ and $f_\pi$ are the neutral pion mass and decay constant, respectively, and we introduced the constant $\Lambda \approx 76\,$MeV. Numerically, Eq.~\eqref{eq:axionmass} translates into the expression $m_0 \approx 5.8{\rm\,\mu eV}(10^{12}{\rm\,GeV}/f_{\rm PQ})$ relating the axion mass $m_0$ with the PQ scale $f_{\rm PQ}$.

\subsection{Cosmological evolution of the axion field}

We consider the evolution of the axion field $\phi$ over a Friedmann-Robertson-Walker metric with line element ${\rm d}s$ squared
\begin{equation}
    {\rm d}s^2 = {\rm d}t^2 - a^2(t)\left({\rm d}r^2 + {\rm d}\vartheta^2 + \sin^2\vartheta\, {\rm d}\varphi^2\right)\,,
\end{equation}
where $a = a(t)$ is the scale factor, and $(t, r,\vartheta,\varphi)$ are the spacetime coordinates. In a cosmological setup which is set below the electroweak scale, the Lagrangian in Eq.~\eqref{eq:Lagrangian} describes the evolution of the axion field across the QCD phase transition. At very high temperatures above the QCD confinement, quarks and gluons combine to form a gas and move freely, in contrast to what they experience at much lower temperatures. When the primordial plasma's temperature drops below the confinement temperature $T_C$, the quark-gluon plasma transitions to the hadronic phase which is characterized by bound quarks forming color-neutral hadrons. The precise value of $T_C$ and the order of the phase transition are subject to intense studies using lattice simulations, see Refs.~\cite{Borsanyi:2016ksw, Petreczky:2016vrs}. The axion field interacts nonperturbatively with the quark-gluon plasma, acquiring the effective temperature-dependent potential given in Eq.~\eqref{eq:Lagrangian} which we parameterize as~\cite{Gross:1980br}
\begin{equation}
    \label{eq:axionpotential}
    V(\phi) = f_{\rm PQ}^2m_\phi^2(T)\left[1-\cos(\phi/f_{\rm PQ})\right]\,,
\end{equation}
where $m_\phi(T)$ encodes the changes in the QCD susceptibility with respect to temperature and $f_{\rm PQ}$ is the PQ symmetry breaking scale. We model the temperature-dependent axion mass as
\begin{equation}
    \label{eq:Tdependentmass}
    m_\phi^2(T) = m_0^2\,\tilde\chi(T)\,,
\end{equation}
where we indicate the axion mass at zero temperature in Eq.~\eqref{eq:axionmass} as $m_0$, $\tilde\chi(T) \equiv \chi(T)/\chi(0)$ and $\chi(T)$ is the QCD susceptibility which we model using the lattice QCD results in Ref.~\cite{Borsanyi:2016ksw}. In this notation, the QCD susceptibility at zero temperature is $\chi(0) = m_0^2f_{\rm PQ}^2 = \Lambda^4$.

Given the Lagrangian in Eq.~\eqref{eq:Lagrangian}, the axion field in the early Universe evolves according to the expression
\begin{equation}
    \label{eq:motion}
    \ddot \phi + 3H\dot\phi + \frac{1}{a^2}\nabla^2\phi + \frac{\partial V}{\partial \phi} = 0\,,
\end{equation}
where a dot is a derivative with respect to cosmic time and $\nabla\phi$ is the spatial gradient of the field $\phi$. On superhorizon scales, the axion field is frozen to its initial configuration $\phi_i = f_{\rm PQ}\theta_i$, where $\theta_i$ is the initial axion angle, until the Hubble rate decreases to the value $H(T) \sim m_\phi(T)$, after which the energy stored in the coherent oscillations of the axion field redshift as a matter field and can be interpreted as the observed dark matter with energy density as
\begin{equation}
    \label{eq:rhophi}
    \rho_\phi = \frac{1}{2}\dot\phi^2 + \frac{1}{2}|\nabla\phi|^2 + V(\phi)\,,
\end{equation}
and the corresponding fractional energy density $\Omega_\phi \equiv \rho_\phi / \rho_{\rm crit}$. The temperature $T_{\rm osc}$ at which the oscillations in the axion field begin is found by solving
\begin{equation}
    \label{eq:Toscdef}
    H(T_{\rm osc}) = m_\phi(T_{\rm osc})\,.  
\end{equation}
We use $T_{\rm osc}$ as a reference temperature around which we start the numerical integration of the axion field equation, as we describe in section~\ref{sec:method}.

\subsection{Axions as dark radiation}
\label{sec:thermalaxions}

Along with the cold axions from the vacuum realignment mechanism, axions as dark radiation are expected to arise from two distinct sources: i) axions emitted by PBH evaporation as Hawking radiation, and ii) thermal axions produced from the decoupling with the primordial plasma at temperature $T_d$. We first discuss the production from the evaporation, for which the contribution from a relativistic species with weighted degrees of freedom $g_{{\rm DR}, H}$ generally corresponds to a deviation in the number of relativistic neutrinos. More in Eq.~\eqref{eq:rhoDR}, or~\cite{Salvio:2013iaa, Baumann:2016wac, Hooper:2019gtx, Caloni:2022uya}
\begin{equation}
    \label{eq:DeltaNeff}
    \Delta N_\nu^{\rm DR} \approx 0.027\, g_{{\rm DR}, H}\,\left(\frac{106.75}{g_S(T_{\rm evap})}\right)^{4/3} \,.
\end{equation}
This value of $\Delta N_\nu^{\rm DR}$ is safely below the current detection limits for all values of $T_{\rm evap}$ of interest here~\cite{Planck:2018vyg}. Now, even if  $N_\nu^{\rm}$  deviates greatly from its standard value, the difference (see Eq.~\eqref{eq:DeltaNeff} for more) will still be within the reach of future CMB probes~\cite{CMB-S4:2016ple, Abazajian:2016hbv, Abazajian:2019eic}, allowing the detection of these light thermal relics~\cite{Baumann:2016wac}.

We now move to the production of thermal axions. The hot QCD axions that get decoupled from the plasma generally have a negligible energy density with respect to the cold component for the mass range that addresses the DM puzzle; in fact, this component is generally neglected in the DM literature.  While its neglected in most of our work, we still - very briefly - assess the said component in PBH-dominated cosmology. However, the computation criterias are not the same as the ones in standard cosmology .\footnote{Previous work derived the abundance of hot axions in a general matter-dominated cosmological era~\cite{Grin:2007yg}.}

For $f_{\rm PQ} \gtrsim 10^{11}\,$GeV, as it is implied in this work, the production of thermal axions is furthered by scattering gluons mediated by the axion-gluon term in the Lagrangian given in Eq.~\eqref{eq:Lagrangian},\footnote{The production of the axion can also take place through processes involving pions~\cite{Chang:1993gm, Hannestad:2005df, Giare:2020vzo} or photons~\cite{Turner:1986tb}.} that leads us to a thermal axion production rate~\cite{Graf:2010tv}
\begin{equation}
    \label{eq:gamma_g}
    \gamma_g \equiv \frac{{\rm d}n_\phi}{{\rm d}t} = \frac{\zeta(3)}{4\pi^5}\,\alpha_s^2\,\frac{T^6}{f_{\rm PQ}^2}\,F_g(T)\,,
\end{equation}
where $\zeta(3) \approx 1.202$ and $F_g(T)$ is a function that adds various contributions to the thermal axion production rate induced by the SM couplings~\cite{Salvio:2013iaa}. With this process, the production of thermal axions proceeds according to a Boltzmann equation for the number density $n_\phi^{\rm th} = n_\phi^{\rm th}(T)$,
\begin{equation}
    \frac{{\rm d}n_\phi^{\rm th}}{{\rm d}t} + 3H(T)n_\phi^{\rm th} = \gamma_g\left(1 - \frac{n_\phi^{\rm th}}{n_\phi^{\rm eq}}\right)\,,
\end{equation}
where $n_\phi^{\rm eq} = (\zeta(3)/\pi^2)T^3$ is the number density at equilibrium. Equivalently, the expression for the yield $Y_\phi^{\rm th} \equiv n_\phi^{\rm th}/s$ reads
\begin{equation}
    \label{eq:thermalaxionsyield}
    \frac{1}{Y_\phi^{\rm th}}\frac{{\rm d}Y_\phi^{\rm th}}{{\rm d}t} = -\frac{\gamma_g}{s} \left(\frac{1}{Y_\phi^{\rm th}} - \frac{1}{Y_\phi^{\rm eq}}\right) + \frac{1}{t_{\rm evap}}\frac{\rho_{\rm BH}}{T s}\,.
\end{equation}
with $Y_\phi^{\rm eq} \equiv n_\phi^{\rm eq}/s = 45 \zeta(3)/(2 \pi^4 g_S(T))$. From Eq.~\eqref{eq:thermalaxionsyield} it is evident that the production of thermal axions and its subsequent dilution by PBH evaporation occurs at two separate instances. The axion component decouples at a temperature $T_d$ given approximately by $\gamma_g(T_d) \sim H(T_d)n_\phi^{\rm th}(T_d)$, or
\begin{equation}
    T_d \approx \frac{8 \pi^4}{3\alpha_s^2} \sqrt{\frac{\pi G_N g_*(T_d)}{5}} \frac{f_{\rm PQ}^2}{F(T_d)} \approx \mathcal{O}(10^{11}{\rm\,GeV}) \,,
\end{equation}
where the numerical result is valid for $f_{\rm PQ} \approx 10^{11}{\rm\,GeV}$. The contribution to $F(T_d) \approx \mathcal{O}(10)$ is mostly given by the top quark and the specific value depends on the model used for the axion-quark couplings~\cite{Salvio:2013iaa}. Any radiation component produced at such a high temperature quickly dilutes away during the PBH domination and gets further diluted by the PBH evaporation. Nevertheless, evaporating PBHs still ejects new hot axions which are further subject to the bound given in Eq.~\eqref{eq:DeltaNeff}.

\subsection{Axion isocurvature fluctuations}
\label{sec:isocurvatures}

Isocurvature modes of a species $i$ with respect to photons are produced as a consequence of a spatial variation in the ratio $n_i/n_\gamma$~\cite{Linde:1985yf, Seckel:1985tj}. If the axion field originates during single-field inflation and the PQ symmetry is not restored afterward, isocurvature modes produced are completely uncorrelated with the curvature perturbations. Here, we consider an axion field produced during a period of inflation occurring at the energy scale $H_I$. The magnitude of axion isocurvature perturbations at a given scale is~\cite{Turner:1990uz, Linde:1991km, Beltran:2006sq}
\begin{equation}
    S_{\rm iso, \phi} = \frac{\Omega_\phi}{\Omega_{\rm DM}}\,\sigma_\phi\,\frac{\partial \ln \Omega_\phi}{\partial \phi}\,,
\end{equation}
where the quantum fluctuations for any nearly-massless scalar field that is present during inflation have variance $\sigma_\phi^2 = H_I^2/(2\pi)^2$ and $\Omega_{\rm DM}$ is the DM abundance. The power spectrum of axion isocurvature fluctuations is then
\begin{eqnarray}
    \Delta_\phi^2(k) &\equiv& \left|S_{\rm iso, \phi}\right|^2 = \left(\frac{\Omega_\phi}{\Omega_{\rm DM}}\right)^2\,\frac{H_I^2}{\pi^2f_{\rm PQ}^2\theta_i^2}\,F(\theta_i)\,,\\
    F(\theta_i) &=& \left(1 + \frac{\partial}{\partial \theta_i}\ln f(\theta_i)\right)^2\,,
\end{eqnarray}
where the function $f(\theta_i)$ accounts for the non-harmonic terms in the axion potential~\cite{Turner:1985si, Strobl:1994wk, Bae:2008ue, Visinelli:2009zm}. More precisely, we assume that the axion energy density can be written as $\Omega_\phi \propto \theta_i^2 f(\theta_i)$.

Results from the {\it Planck} Collaboration quote the bound on the fraction of uncorrelated (axion) isocurvature fluctuations
\begin{equation}
    \beta(k_0) \equiv \frac{\Delta_\phi^2(k_0)}{\Delta_\phi^2(k_0) + \Delta_{\mathcal{R}}^2(k_0)}\,,
\end{equation}
where $\Delta_{\mathcal{R}}^2(k)$ is the curvature power spectrum with wavenumber $k$ and $k_0$ is a pivotal scale. Using the CMB temperature and polarisation anisotropy data jointly with the BICEP2/Keck Array at the scale $k_0 = 0.002{\rm\,Mpc}^{-1}$, the joint collaborations report the bound $\beta(k_0) \lesssim 0.035$ at 95\% confidence level~\cite{Planck:2018jri} and the measurement of the power spectrum $A_s = 3.044 \pm 0.014$ at 68\% confidence level, where $A_s \equiv \ln(10^{10}\Delta_{\mathcal{R}}^2(k_0))$~\cite{Planck:2018vyg}. For scenarios with axion being the dominant contribution to the DM and for $\theta_i\lesssim \pi$, the bound translates into
\begin{equation}
    \label{eq:isocurvatures}
    H_I \lesssim 3\times 10^{-5}\,f_{\rm PQ}\,\theta_i/F(\theta_i)\,.    
\end{equation}
Note, this expression holds in the standard cosmology as well as in the non-standard formulations, although the value of $f_{\rm PQ}$ and eventually the expression for $f(\theta_i)$ would be modified.

The difficulty in building these types of models lies in the prerequisites that include a low energy scale of inflation from isocurvature bounds. Explicit models of low-scale inflation have been suggested to accommodate the axion in such a scenario~\cite{Graham:2018jyp, Takahashi:2018tdu, Schmitz:2018nhb, Tenkanen:2019xzn}. In models where the axion field is generated prior to inflation, we do not observe any topological defects as inflation washes out all such inhomogeneities (which, further, also includes axion strings). Other scenarios with plummeting tensor-to-scalar ratios invoke the usage of $f(R)$ gravity models with a non-minimal coupling to the axion field~\cite{Odintsov:2019mlf, Odintsov:2019evb, Lambiase:2022ucu}.

\section{Method}
\label{sec:method}

We describe the numerical methods used to derive the results in our work. The accompanying material can be found here:~\href{https://github.com/lucavisinelli/AXIONPBH}{github.com/lucavisinelli/AXIONPBH}, and the corresponding set of equations is discussed in Appendix~\ref{sec:axionEOM}. The goal of the code is to provide the mass of the axion that accounts for the present DM abundance given the details of the cosmology in the form of $(\beta, M_{\rm BH})$ and the initial axion angle $\theta_i$. For this, the expression for PBH abundances in Eq.~\eqref{eq:rhovsT} is solved with the initial condition in Eq.~\eqref{eq:definebeta} as a function of temperature over the vast range $(T_f, T_{\rm BBN})$, once the independent variable is chosen as $\tau \equiv \ln(T/T_f)$.

We now address the computation of the axion abundance. For a given axion mass $m_0$, we first find the temperature $T_{\rm osc}$ that serves as a proxy for the axion oscillations by solving Eq.~\eqref{eq:Toscdef}, with the Hubble rate given in terms of the numerical solution of Eq.~\eqref{eq:rhovsT} and the QCD susceptibility from Ref.~\cite{Borsanyi:2016ksw}. After the onset of field oscillations, the axion field evolves as a cold component with a number density $n_\phi \equiv \rho_\phi/m_\phi(T)$ which is a solution of
\begin{equation}
    \frac{{\rm d}n_\phi}{{\rm d}t} + 3H n_\phi = 0\,.
\end{equation}
To account for the additional dilution resulting from PBH evaporation, we track the change in the axion yield $Y_\phi \equiv n_\phi/s$, where the entropy density follows Eq.~\eqref{eq:entropy}, and we get,
\begin{equation}
    \label{eq:Y}
    \frac{1}{Y_\phi}\frac{{\rm d}Y_\phi}{{\rm d}t} = \frac{{\rm d}M_{\rm BH}/{\rm d}t}{M_{\rm BH}}\,\frac{\rho_{\rm BH}}{T s}\,,
\end{equation}
which now allows us to evaluate the value of the quantity $\ln [Y_\phi / Y_\phi^{\rm osc}]$ across the PBH evaporation epoch. In practice, we integrate Eq.~\eqref{eq:Y} from temperature T $\gtrsim T_{\rm osc}$ to $T_{\rm BBN}$, at which we stop the numerical integration. The most resource-consuming aspect of the resolution lies in obtaining the initial value of the yield, namely the quantity $Y_\phi^{\rm osc} = \rho_\phi^{\rm osc}/(m(T_{\rm osc})\,s(T_{\rm osc}))$. For this, we numerically solve the axion equation of motion in Eq.~\eqref{eq:motion} for the initial value $\theta_i$ and for the superhorizon modes $|\nabla\phi|\ll a$, in terms of the independent variable $\xi \equiv \ln(T/T_{\rm osc}) = \tau + \ln(T_f/T_{\rm osc})$ which greatly simplifies the scheme.\footnote{See Ref.~\cite{Hoof:2018ieb} for an alternative expression of Eq.~\eqref{eq:motion} in terms of the temperature $T$.}

The value of the yield $Y_\phi$ obtained at temperature $T_{\rm BBN}$ has remained constant to date. The relic abundance of DM axions is then 
\begin{equation}
    \label{eq:axionDM}
    \Omega_\phi(T_0) = \frac{m_0}{m_\phi(T_{\rm BBN})}\, \frac{g_S(T_0)T_0^3}{g_S(T_{\rm BBN})T_{\rm BBN}^3}\,\frac{\rho_\phi(T_{\rm BBN})}{\rho_{\rm crit}(T_0)}\,,
\end{equation}
where $T_0$ is the present CMB temperature and $\rho_{\rm crit}(T_0) = 3H_0^2/(8\pi G_N)$ is the present critical density as a function of the Hubble constant $H_0$. The present number of entropy degrees of freedom is $g_S(T_0) \approx 3.909$ while before neutrino decoupling it is $g_S(T) \approx g_*(T)$. Because value of the axion mass $m_0$ does not guarantee that the abundance given in Eq.~\eqref{eq:axionDM} corresponds to the observed dark matter, we iterate over $m_0$ until the quantity $|\Omega_\phi(T_0)/\Omega_{\rm DM} - 1|$ falls below the desired tolerance, here $\epsilon = 10^{-3}$.

\section{Results}
\label{sec:results}

Using the numerical tools developed in section~\ref{sec:method}, we compute the axion mass as a function of the PBH parameter space and initial axion angle. Safe to say that the results impact the cosmological model in question.  We consider PBHs in the mass range 
\begin{equation}
    10^6{\rm\,g} \lesssim M_{\rm BH} \lesssim 5\times 10^8{\rm\,g}\,,
\end{equation}
where the lower limit is obtained from setting $T_{\rm osc} = T_{\rm EQ}$ and has a dependence on $\beta$. A value in the ballpark of $10^6{\rm\,g}$ is deduced by demanding that PBHs do not evaporate at temperatures $T_{\rm osc} \sim 1\,$GeV. This temperature also corresponds to when the axion field begins to oscillate. The upper limit is deciphered from the fact that PBHs disappeared before BBN; see Eq.~\eqref{eq:BBNconstraint}.

\begin{figure}
    \centering
    \includegraphics[width=0.8\linewidth]{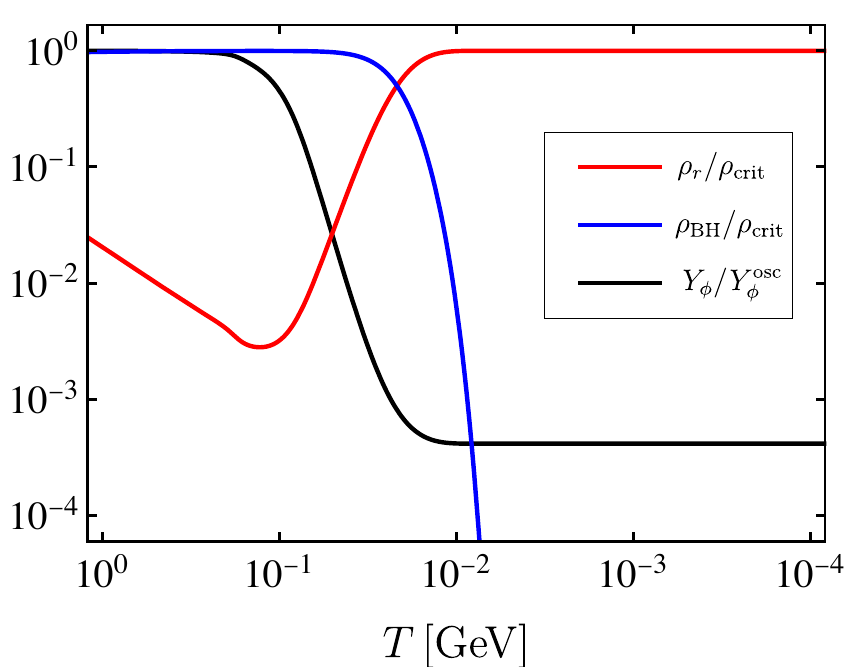}
    \caption{The black line shows the ratio of the axion yield $Y_\phi$ to its value at the onset of oscillations $Y_\phi^{\rm osc}$, as a function of the plasma temperature $T$. We set the initial PBH abundance $\beta = 10^{-10}$ with a monochromatic mass distribution centered at $M_{\rm BH} = 10^8\,$g. Also shown are the relative abundances of the energy densities in PBHs (blue line) and radiation (red line).}
    \label{fig:yield}   
\end{figure}

Figure~\ref{fig:yield} shows the yield $Y_\phi$ obtained by solving Eq.~\eqref{eq:Y} (black line) and in units of $Y_\phi^{\rm osc}$, for the choice of the initial PBH abundance $\beta = 10^{-10}$ with a monochromatic mass distribution centered at $M_{\rm BH} = 10^8\,$g. For this set of parameters, PBHs have come to dominate the expansion rate by the time the axion field begins to oscillate to then evaporate at $T_{\rm evap}\approx 20\rm\,$MeV, see the blue line in figure~\ref{fig:yield}. The fractional energy density in radiation is the red line in the same figure. Note, prior to PBH evaporation, the yield decreases from its initial value $Y_\phi^{\rm osc}$ because of the additional entropy injected by the particles emitted as Hawking radiation. This aspect is peculiar to the PBH-dominated epoch and its effect has to be taken into account to compute the abundance of DM axions today in a proper manner~\cite{Visinelli:2009kt}.

\begin{figure}
    \centering
    \includegraphics[width=0.8\linewidth]{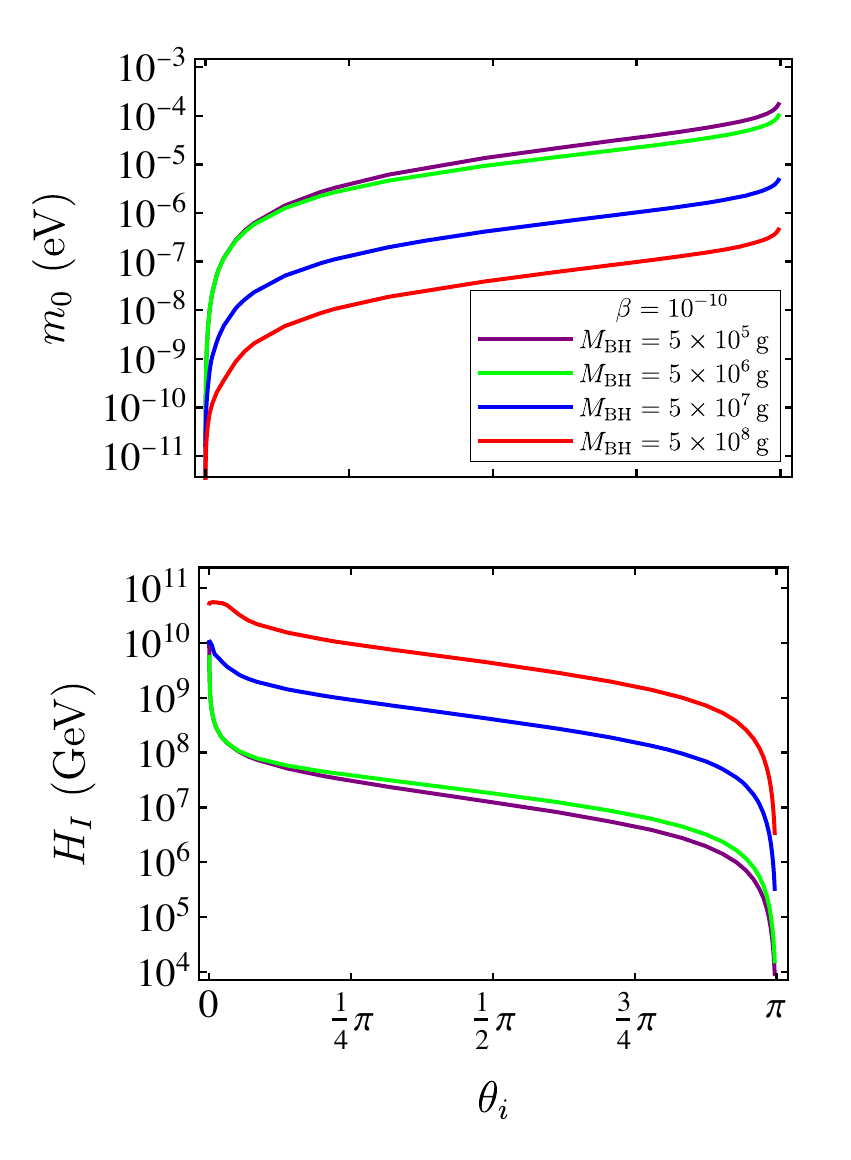}
    \caption{{\it Top panel}: The DM axion mass $m_0$ in eV as a function of the initial axion angle $\theta_i$, for the initial fractional PBH abundance $\beta = 10^{-10}$ and for different choices of the PBH mass: $M_{\rm BH} = 5\times 10^5\,$g (black line), $M_{\rm BH} = 5\times 10^6\,$g (green line), $M_{\rm BH} = 5\times 10^7\,$g (blue line), and $M_{\rm BH} = 5\times 10^8\,$g (red line). {\it Bottom panel}: The lower bound on the inflation scale $H_I$ in Eq.~\eqref{eq:isocurvatures} as a function of the initial axion angle $\theta_i$. The color code is the same as the top panel.}
    \label{fig:Plotmassvstheta}
\end{figure}

We now include differing values of initial axion angle $\theta_i$ and study its impact. In the top panel of figure~\ref{fig:Plotmassvstheta} we report the relation between the DM axion mass in eV and the initial axion angle $\theta_i$, for $\beta = 10^{-10}$ and for different PBH masses $M_{\rm BH}$, namely $M_{\rm BH} = 5\times 10^5\,$g (black line), $M_{\rm BH} = 5\times 10^6\,$g (green line), $M_{\rm BH} = 5\times 10^7\,$g (blue line), and $M_{\rm BH} = 5\times 10^8\,$g (red line). We first discuss the decrease in $m_0$ with increasing values of $M_{\rm BH}$, for a fixed $\theta_i$. Because of the extra dilution from entropy injection, the axion energy density in the early PBH domination epoch is generally lower than the one in standard cosmology for a given axion mass. The present energy density, being usually negatively dependent on $m_0$ in both standard and early matter-dominated cosmologies, demands axions to correspond to DM with lower values of $m_0$. We now comment on the increase of the axion mass with $\theta_i$ for a given value of $M_{\rm BH}$, which comes from the positive dependence of the energy density on $\theta_i$ and negative dependence on $m_0$, and both the axion angle and its mass must increase in order to attain the value of the present energy density. PBHs of mass $M_{\rm BH} = 5\times 10^5\,$g  would not affect the dynamics of the axion since they would have fully evaporated before the axion field began oscillating which makes this scenario indistinguishable from a standard one. For all other scenarios, the axion mass increases drastically with $\theta_i \approx \pi$, because of the non-harmonic terms in the potential. Moreover, given a value for $\theta_i$, it is generally expected that the DM axion mass decreases with increasing PBH mass since a larger $M_{\rm BH}$ leads to a longer period of PBH domination, thus enhancing the effects of entropy dilution.

The bottom panel of figure~\ref{fig:Plotmassvstheta} gives the bound on the inflation scale $H_I$ from Eq.~\eqref{eq:isocurvatures}, so that the region above each curve is excluded for the specific choice of cosmological model. As is well known from literature~\cite{Turner:1990uz, Linde:1991km, Beltran:2006sq}, the bound on the isocurvature modes leads to a stringent constraint on $H_I$, which for the PQ scale $f_{\rm PQ}\sim 10^{12}\,$GeV corresponds to $H_I \lesssim 2\times 10^7\,$GeV. While models of low-scale inflation exist~\cite{Graham:2018jyp, Takahashi:2018tdu, Schmitz:2018nhb, Tenkanen:2019xzn}, single-field inflation with an approximately quadratic potential gives $H_I \sim 10^{12}\,$GeV once the requirements for the amplitude of the perturbations in the CMB are satisfied. This can be reconciled by considering a smaller value of $\theta_i$ or, as shown in figure~\ref{fig:Plotmassvstheta}, by incorporating the implicit effects of the PBH population on the prerequisites for a higher PQ scale that lowers the isocurvature power spectrum.

Note, for $M_{\rm BH} \gtrsim 5\times 10^7\,$g and for small values of $\theta_i$, the bound in the bottom panel of figure~\ref{fig:Plotmassvstheta} seems to reach a plateau which is not present for smaller PBH masses. This does not result from a numerical error, it actually comes from the fact that when the axion field in question begins to oscillate at temperatures below the QCD phase transition, its mass is no longer dependent on temperature. The PQ scale for the DM axion is, therefore, $f_{\rm PQ} \propto \theta_i^{-1}$, see Eq.~(83) in Ref.~\cite{Visinelli:2009kt}. In this regime, the bound in Eq.~\eqref{eq:isocurvatures} becomes independent of $\theta_i$. At the same time, for lower values of the PBH mass the PQ scale is $f_{\rm PQ} \propto \theta_i^{-4/3}$, leading to a mild dependence of the bound on $H_I$ on the initial axion angle.

\begin{figure}
    \centering
    \includegraphics[width=1.0\linewidth]{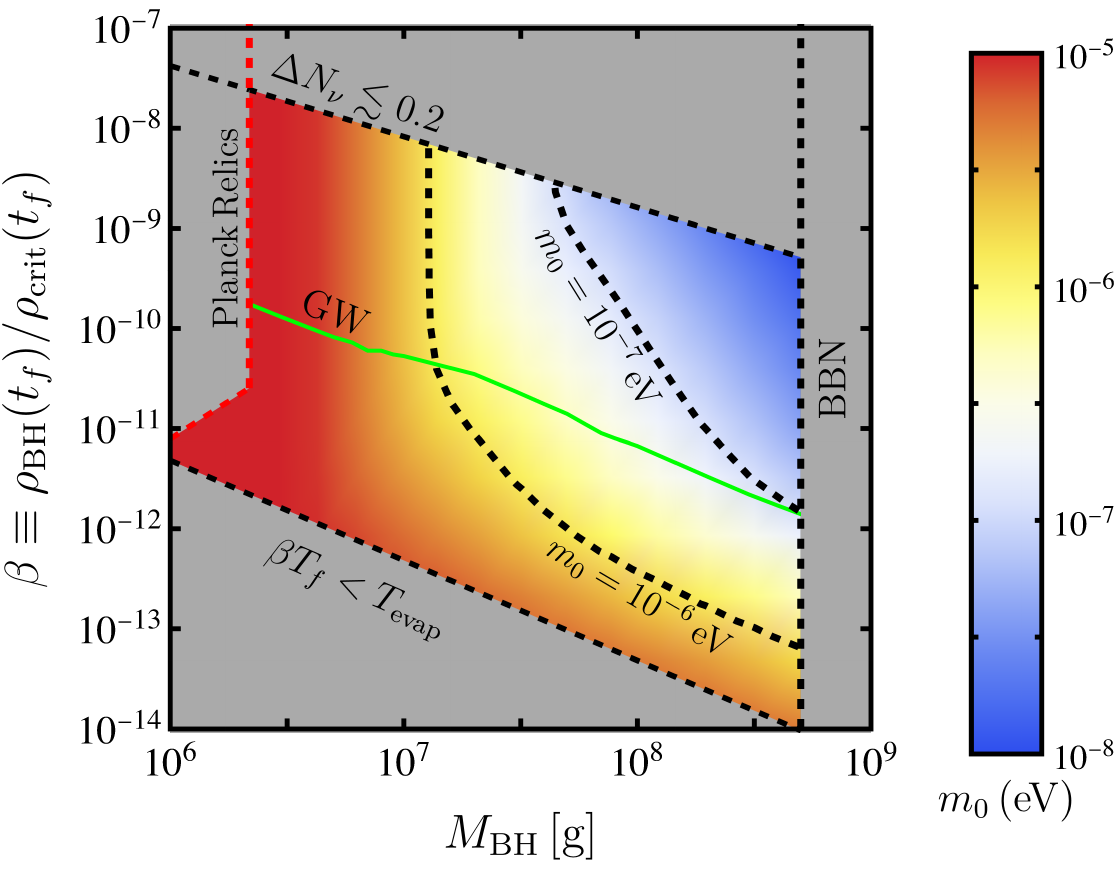}
    \caption{The mass of the DM axion $m_0$ in eV for different PBH initial fractional abundances $\beta$ (vertical axis) and PBH mass $M_{\rm BH}$ in grams (horizontal axis), with the initial axion angle $\theta_i = 1$. We have marked some of the specific values of the DM axion mass with dashed black lines. Also shown is the bound on the PBH mass in Eq.~\eqref{eq:BBNconstraint} labeled ``BBN'', the upper bound in Eq.~\eqref{eq:boundbeta} labeled ``$\Delta N_\nu \lesssim 0.2$'', and the lower bound in Eq.~\eqref{eq:lowerbound1} and labeled ``$\beta T_f < T_{\rm evap}$''. The area bounded by the dashed red line and labeled ``Planck relics'' comes possibly from an incomplete PBH evaporation process. The green curve labeled ``GW'' is the forecast of the upper bound on the isocurvature-induced GWs that are within reach once future GW detectors (DECIGO, ET, LISA) will go online, see text for details.}
    \label{fig:densityplot}
\end{figure}

We now shift our focus to assessing the axion mass as a function of the PBH parameter space $(\beta, M_{\rm BH})$. The color code in the density plot in figure~\ref{fig:densityplot} returns the mass that ensures the axion is the candidate for DM in the cosmological model described by the set $(\beta, M_{\rm BH})$. We have fixed the initial axion angle $\theta_i = 1$ for when the oscillations begin, and we have used the numerical code described in section~\ref{sec:method} to derive the results on a grid of points for PBH parameters. For masses $M_{\rm BH} \lesssim 5\times 10^6\,$g, PBHs evaporate before the axion field begins to oscillate and the DM axion mass remains unchanged with respect to the standard cosmological scenario. The DM axion mass shifts to smaller values for higher $M_{\rm BH}$, for which the duration of PBH domination is increased as the evaporation is delayed, and for higher values of $\beta$ for which PBH domination starts earlier, see figure~\ref{fig:BKGD}. This occurs because of the inverse proportionality between the axion energy density and its mass, so that in order to counteract the entropy dilution of the energy density, the value of $m_0$ needs to be decreased to obtain the present DM abundance in all cosmological models. The smallest achievable value of the DM axion mass that is consistent with the bound is $m_0 \sim 10^{-8}\,$eV. However, this value may further go lower if $\theta_i$ is decreased, as the red curve in figure~\ref{fig:Plotmassvstheta} indicates.

In figure~\ref{fig:densityplot}, the bound labeled ``$\Delta N_\nu \lesssim 0.2$'' has been tacitly derived in Ref.~\cite{Domenech:2020ssp} (see their figure~4), by considering the GW energy at recombination as an additional contribution to $N_\nu$, as discussed in section~\ref{sec:PBHisobounds}. Two considerations are in place. First, the bound will improve in the near future once more stringent measurements of $\Delta N_\nu$ are available. For instance, the CMB Stage IV experiments forecast $\Delta N_\nu \lesssim 0.06$ at 95\% CL~\cite{CMB-S4:2016ple, Abazajian:2016hbv, Abazajian:2019eic}. Second, future detectors will be able to detect the primordial GW released in the model directly. In fact, the GW frequency is expected to peak at around $(0.1 - 10)\,$Hz for the PBH mass range considered, which coincides with the range that will be accessible with DECIGO~\cite{Seto:2001qf, Yagi:2011wg} and the Einstein Telescope (ET)~\cite{Maggiore:2019uih}, and partially also by LISA~\cite{2017arXiv170200786A}. Using the formulation leading to figure~3 in Ref.~\cite{Domenech:2020ssp} and the expected sensitivity of the future GW detectors, we have derived the region of the parameter space which will be in reach of these GW detectors once they will be online, lying above the green line in figure~\ref{fig:densityplot} labeled as ``GW''. In more detail, for a given value of $M_{\rm BH}$, the green line gives the minimum value of $\beta$ for which the GW amplitude expected from the model appears in one of the future detectors.\footnote{A weaker upper bound resulting from the consequences of PBH evaporation on the CMB, has been derived in Ref.~\cite{1977SvAL....3..110Z}.} The area bounded by the dashed red line and labeled ``Planck relics'' is obtained by assuming that PBHs do not evaporate completely, leaving a relic of mass $G_N^{-1/2}$ which adds up to the DM budget~\cite{Carr:1994ar}. The lower bound labeled ``$\beta T_f < T_{\rm evap}$'' is obtained from the consistency requirement that PBHs come to dominate the content before their evaporation and prior to BBN, see Eq.~\eqref{eq:lowerbound1}. See figure~6 in Ref.~\cite{Bernal:2021bbv} for additional bounds coming from the assessment against the forecast reach in CMB Stage IV experiments.

\section{Discussions}
\label{sec:discussion}

The aforementioned scenario is bound to have a pronounced impact on several observables, making the results obtained testable to a certain extent. We worked under the assumption that the PQ symmetry was already broken during inflation, effectively washing out the unobserved topological defects that inevitably result from the spontaneous breaking of the PQ symmetry. Still, most of the results obtained also hold in the opposite regime where inflation does not help with getting rid of the said topological defects and the PQ symmetry is broken (or restored) at a lower energy scale~\cite{Bernal:2021bbv, Bernal:2021yyb}. In the latter case, both the non-relativistic and relativistic populations benefit from the additional injection of entropy coming from the decay of axionic strings, whose net effect modifies the DM axion mass by a factor $\sim \mathcal{O}(1 - 10)$.

\subsection{Impact on axion substructures}
\label{sec:miniclusters}

The delayed onset of axionic field oscillations in the early matter-dominated epoch leads to heavier substructures such as axion miniclusters~\cite{Hogan:1988mp}, since a smaller $T_{\rm osc}$ leads to a higher horizon length scale~\cite{Visinelli:2018wza, Nelson:2018via, Blinov:2019jqc}. In addition, the model predicts an early stage of cosmic structure formation~\cite{Erickcek:2011us}, which might increase the typical minicluster mass~\cite{Visinelli:2018wza}. Despite the recent progress in the simulation of the axionic string network~\cite{Klaer:2017ond, Gorghetto:2018myk, Vaquero:2018tib, Gorghetto:2020qws, Buschmann:2021sdq, Eggemeier:2022hqa}, a dedicated study involving an early period of matter domination is lacking, but see Ref.~\cite{Yamaguchi:1999dy} for an earlier study along this line. Assuming that the DM axion mass shifts by $\mathcal{O}(1)$ due to topological defects, the picture in figure~\ref{fig:densityplot} remains valid and a heavier PBH mass would correspond to a lighter DM axion. Note, that in some scenarios the changes due to topological defects can be sensibly larger and the impact on the axion mass can be a factor of $\mathcal{O}(10)$ or larger~\cite{Gorghetto:2020qws}. Both the existence of axion miniclusters and the length scale affected, including the change brought in by the presence of the early PBH domination, may be exploited by future pulsar timing experiments~\cite{Lee:2020wfn} as the window associated with measuring the time period of pulsars is sensitive to sub-halos of mass $M \gtrsim 10^{-13}M_\odot$~\cite{Dror:2019twh}.

Note, that pulsar timing measurements are subject to the late-Universe abundance of miniclusters and other axion substructures, which have to be adequately modeled within the Milky Way to extract useful information. Considerable efforts in this direction has been taken in recent year pertaining to the effect of tidal disruption of miniclusters from nearby stars and the galactic gravitational field~\cite{Dokuchaev:2017psd, Kavanagh:2020gcy, Edwards:2020afl, Shen:2022ltx} as well as the reconstruction of the fraction of dark matter axions bound in miniclusters~\cite{Eggemeier:2022hqa}. A possible alternative for this scenario that is less impacted by galactic disruption phenomena is lensing~\cite{Dai:2019lud}.

Besides pulsar timing arrays, other experiments will hopefully be able to test these scenarios in the coming decades. Novel techniques are sparking a mounting interest in the detection of axions in the low-mass range. One such possibility involves creating a superconducting cavity which can comprehend data as low as the one associated with axion-photon coupling $g_{a\gamma} \sim 10^{-18}\, \mathrm{GeV}^{-1}$~\cite{Berlin:2019ahk, Berlin:2020vrk, Berlin:2022hfx}, enabling the detection of the DM axions with a mass in the $\mathcal{O}(10)\,$neV range. This mass range overlaps with the predictions of the proposed model, so that detection of the axion in a future superradiant cavity could pave a path toward experimentally proving the existence of a primordial population of BHs that has been illustrated in figure~\ref{fig:densityplot}.

These assumptions need to be cross-checked with the potential reaches that will, in turn, be obtained by the CMB Stage IV arrays, for which another independent probe could be through GWs~\cite{Domenech:2020ssp}. The extended array of LISA~\cite{2017arXiv170200786A}, DECIGO~\cite{Seto:2001qf, Yagi:2011wg}, and the Einstein Telescope~\cite{Maggiore:2019uih} would be able to detect the GW packet associated with relatively higher values of $\beta$ within the parameter space allowed in figure~\ref{fig:densityplot}.

As we discussed before, the shift in the axion mass leads to a change in the peak mass associated with the distribution of miniclusters. Such effects can be used to test the presence of an early matter-dominated era with respect to other effects that could have been related to a shift in the axion mass towards lighter values in the early Universe. As discussed in Ref.~\cite{Visinelli:2018wza}, the minicluster mass and radius are affected by the presence of a matter-dominated cosmology in two ways: i) the modified cosmology leads to a lighter axion and to a delay in axion oscillations, at times where the Hubble radius is larger; ii) Some degree of structure formation already present at that time could impact the minicluster mass by several orders of magnitude. Both these effects shift the characteristic minicluster mass towards heavier values. Besides the peak mass, another signature is the velocity dispersion of the axions in the minicluster; which also affects the cutoff of the low-end mass spectrum. Overall, a shift in the minicluster mass spectrum would be a sign that an epoch of early matter domination indeed occurred.

\subsection{Extended mass distribution}
\label{sec:extendedmassdistribution}

We now discuss how realistic the assumption of a monochromatic mass function is, along with the implications that may follow by dropping such a requirement. In fact, even in the presence of a mechanism that leads to a peak BH mass, PBHs are not only produced with horizon mass $M_H$ at the time of formation but also with much smaller masses, according to a scaling formula~\cite{Niemeyer:1997mt}
\begin{equation}
    M = K M_H(\delta-\delta_c)^\eta\,,
\end{equation}
which is valid for all overdensities $\delta$ above the critical value $\delta_c \approx 0.4\textrm{--}0.7$~\cite{Shibata:1999zs, Niemeyer:1999ak, Musco:2004ak} and $K$, $\eta$ are parameters with $\eta \approx 0.36$~\cite{Koike:1995jm, Niemeyer:1999ak} and $K \approx 3.3$~\cite{Niemeyer:1997mt}. The fraction of the energy density in collapsing PBHs at a given epoch as a function of the root-mean-square fluctuation amplitude $\sigma$ of the primordial power spectrum of density perturbations is~\cite{Carr:1975qj, Green:2004wb}
\begin{equation}
    \beta \approx K\sigma^{2\eta}\,{\rm erfc}\left(-\frac{\delta_c}{\sqrt{2}\sigma}\right) \approx K\sqrt{\frac{2}{\pi}}\frac{\sigma^{2\eta+1}}{\delta_c}\,\exp\left(-\frac{\delta_c^2}{2\sigma^2}\right)\,,
\end{equation}
which is valid for $\sigma \ll \delta_c$. It is then generally expected that a monochromatic mass distribution is an approximation at best. Demanding that $\beta$ lies within the range $[10^{-14} - 10^{-8}]$ obtained in figure~\ref{fig:densityplot} gives the range $\sigma = [0.13 - 0.17]\delta_c$. These values can be related to the shape of the primordial potential at the time of PBH formation~\cite{Green:2004wb}.

An extended PBH distribution would generally be described by a larger set of parameters that account for the formation mechanism of PBHs over an extended period of time and the distortion caused by astrophysical effects such as merging and accretion, see section~\ref{sec:accretion}. For a PBH distribution described by a mass function ${\rm d}n/{\rm d} \ln M_{\rm BH}$, the primordial abundance is
\begin{equation}
    \beta = \frac{1}{\rho_{\rm crit}(t_f)}\int {\rm d}M_{\rm BH}\,\frac{{\rm d}n}{{\rm d} \ln M_{\rm BH}}\,,
\end{equation}
where the energy density $\rho_{\rm crit}(t_f)$ is associated to the collapse of a region of mass $M_f$. An example that generalizes the previous results is the log-normal mass distribution which describes a large class of inflationary PBH models~\cite{Green:2016xgy},
\begin{equation}
    \frac{{\rm d}n}{{\rm d}\ln M_{\rm BH}} \propto \exp\left[-\frac{(\ln M_{\rm BH}/M_f)^2}{2\sigma_f^2}\right]\,,
\end{equation}
where $M_f$ is the peak mass of the distribution with variance $\sigma_f^2$. The use of extended distributions is expected in specific models of PBH formation, in light of recent claims about the difficulty to obtain a monochromatic spectrum in single-field inflation~\cite{Kristiano:2022maq, Inomata:2022yte}. Another factor that severely influences the outcome of PBH formation is the presence of non-Gaussianities in the distribution of overdensities~\cite{Young:2013oia, Young:2015kda}, which has been extensively treated in the literature (see Ref.~\cite{Green:2020jor} for a review).

\subsection{Comparison with moduli field decay}
\label{sec:modulifield}

So far, we have discussed the assessment of the dark matter axion density in a PBH-dominated cosmology. Various different models have previously been proposed in which the Universe is filled by a matter-like field for some period  prior to BBN, including the dominance of moduli fields~\cite{Moroi:1999zb, Gelmini:2006pq, Gelmini:2008sh}, the post-inflation reheating stage~\cite{Chung:1998rq, Chung:1998zb}, or thermal inflation~\cite{Lyth:1995ka}. For instance, a model in which a non-relativistic moduli field $\Psi$ predominantly decays into relativistic particles with a branching ratio $\simeq 1$, the relative abundance of these components is regulated by the expression~\cite{Moroi:1999zb}
\begin{eqnarray}
    \frac{{\rm d}\rho_\Psi}{{\rm d}t} + 3H\rho_\Psi &=& -\Gamma \rho_\Psi\,,\\
    \frac{{\rm d}\rho_r}{{\rm d}t} + 4H\rho_r &=& \Gamma \rho_\Psi\,,
\end{eqnarray}
where $\rho_\Psi$ is the energy density in the moduli field, of mass $m_\Psi$ and decay width
\begin{equation}
    \label{eq:modulidecayrate}
    \Gamma = \frac{1}{4\pi}\frac{m_\Psi^3}{m_{\rm Pl}^2}\,.
\end{equation}
The main difference of this scenario with respect to what has been presented in section~\ref{sec:evaporation} consists in the time-dependence of the PBH decay rate, given by the mass rate change. Here, we have neglected such a change, and we have checked numerically that the approximation is justified for the model considered, so that the two scenarios of moduli or PBH domination are formally identical.

A closer look into the decay formulae reveals the desired parameter space for the two models so that the decay occurs before BBN. Given the reheating temperature $T_{\rm RH}$, at which the Universe transitions into the standard cosmology, the decay rate of the massive component has to match $\Gamma \sim T_{\rm RH}^2/m_{\rm Pl} \gtrsim 1/$s. We then obtain
\begin{equation}
   \frac{1}{t_{\rm evap}} \approx \frac{1}{0.2{\rm \, s}}\,\left(\frac{M_{\rm BH}}{5\times 10^8{\rm\,g}}\right)^3\,,
\end{equation}
while the expression in Eq.~\eqref{eq:modulidecayrate} for a moduli field gives
\begin{equation}
   \Gamma \approx \frac{1}{1.2{\rm\,s}}\,\left(\frac{m_\Psi}{100{\rm\,TeV}}\right)^3\,.
\end{equation}

We now assess the effect of a mass-dependent decay rate from a numerical resolution of the corresponding Boltzmann equation. We consider a model in which the temperature dependence of the degrees of freedom and of the axion field are neglected, in order to underline the effects due to the mass change that we generalize with respect to Eq.~\eqref{eq:masslossrate} as
\begin{equation}
    \label{eq:decayrate}
    \frac{{\rm d}M_{\rm BH}/{\rm d}t}{M_{\rm BH}} = - \frac{\alpha}{(M_{\rm BH})^b}\,,
\end{equation}
where $\alpha$ is a constant and $b$ parametrizes the rate of the mass change. For example, $b = -1$ for a constant rate such as the moduli field scenario and $b = 2$ for the Hawking mechanism. We first focus on the case $b \neq -1$. The evolution of the PBH mass is given by
\begin{equation}
    M(t) = M_0 \left(1 - t/t_{\rm evap}\right)^{1/(1+b)}\,,
\end{equation}
where the initial mass is $M(0) = M_0$ and $t_{\rm evap} \equiv M_0^{1+b}/(1+b)/\alpha$. For $b = -1$, we instead obtain $M(t) = M_0\exp(-\alpha_{-1} t)$, where $\alpha_{-1}$ is the value of $\alpha$ for the case of a constant decay rate, which is fixed by equating the decay rate for the two different cases at $t = 0$ as $(1-b)\alpha_{-1}t_{\rm evap} = 1$. The set of Eqs.~\eqref{eq:PBHmassloss}--\eqref{eq:rad} is then rewritten in terms of $x = t/t_{\rm evap}$ as
\begin{eqnarray}
    \frac{{\rm d}y}{{\rm d}x} + (y+z)^{1/2}y &=& -\frac{y}{1 - \kappa x}\,,\label{eq:y}\\
    \frac{{\rm d}z}{{\rm d}x} + \frac{4}{3}(y+z)^{1/2}z &=& \frac{y}{1 - \kappa x}\,,\label{eq:z}
\end{eqnarray}
with $\kappa = 0$ for $b = -1$ and $\kappa = 1$ otherwise, and where we have rescaled $\rho_{\rm BH}$ and $\rho_r$ with the quantity $\bar\rho \equiv (24\pi G_N (1+b)^2t_{\rm evap}^2)^{-1}$ to obtain $y$ and $z$, respectively.

\begin{figure}
    \centering
    \includegraphics[width=0.8\linewidth]{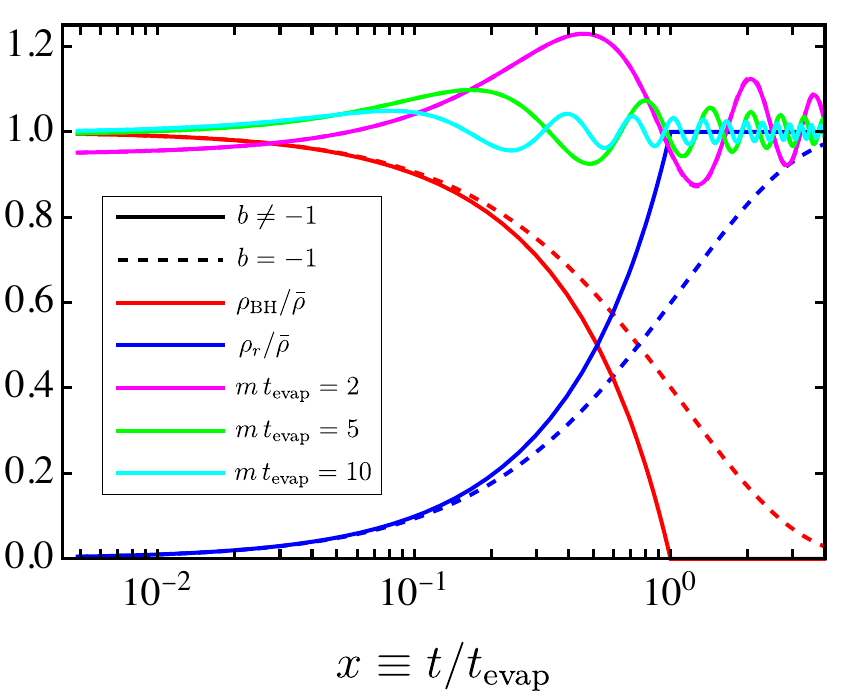}
    \caption{The solution to the background Eqs.~\eqref{eq:y}--\eqref{eq:z} for a varying decay rate with $b \neq -1$ (solid lines) and for a constant decay rate with $b = -1$ (dashed lines), see Eq.~\eqref{eq:decayrate} for the definitions. Lines are shown for the PBH (red) and radiation (blue) energy densities, as a function of the dimensionless time $x \equiv t/t_{\rm evap}$. Also shown is the normalized axion density times the Hubble volume $\rho_\phi a^3$ for different values of the axion mass: $mt_{\rm evap} = 2$ (magenta), $mt_{\rm evap} = 5$ (green), and $mt_{\rm evap} = 10$ (cyan).}
    \label{fig:ALPs}   
\end{figure}

In the setup above, we derived the abundance of an axion field $\phi$ propagating according to Eq.~\eqref{eq:motion} with a quadratic potential and with a constant mass $m$, so that the equation for the axion field with respect to the rescaled time coordinate $x$ becomes
\begin{equation}
    \label{eq:phi}
    \frac{{\rm d}^2\phi}{{\rm d}x^2} + (y+z)^{1/2}\frac{{\rm d}\phi}{{\rm d}x} + (m\,t_{\rm evap})^2 \phi = 0\,.
\end{equation}
In figure~\ref{fig:ALPs} we show the background obtained from Eqs.~\eqref{eq:y}--\eqref{eq:z} for the PBH density (red) and for the radiation energy density (blue) and for the case of a varying decay rate with $b \neq -1$ (solid lines) or a constant decay rate with $b = -1$ (dashed line), as well as the quantity $\rho_\phi a^3$ derived from Eq.~\eqref{eq:phi} and normalized with respect to its average over various oscillations. The axion energy density is shown for $mt_{\rm evap} = 2$ (magenta), $mt_{\rm evap} = 5$ (green), and $mt_{\rm evap} = 10$ (cyan). Some considerations are in place. The moment of PBH-radiation equality occurs at a time $x_{\rm EQ}$ which differs by a factor $\mathcal{O}(1)$ between the two models, making the thermal histories in the two scenarios coincident. This is reflected in the various solutions obtained for the axion energy density, for which there is no appreciable difference between the dashed and solid lines for each value of the axion mass. This is confirmed by the average value over multiple oscillations, which is modified by the effect of a varying decay rate (by $\lesssim 1\%$).

\section{Conclusions}
\label{sec:conclusions}

We investigated how a population of primordial black holes that briefly came to dominate the expansion rate of the Universe, prior to BBN, affect the present abundance of the QCD axion. In the model, the mass distribution of the PBH population is a monochromatic mass spectrum that peaked at a value smaller than $\approx 5\times 10^{8}\,$g, leading them to evaporate completely prior to BBN so it does not end up affecting the light element abundances. We primarily studied the layout where the population of PBHs dominates the energy density of the early Universe and how the said population goes on to grow over time and contributes a much more significant part to the total energy density.

Assuming that the Peccei-Quinn symmetry is broken during inflation, we have discussed how a modified cosmology affects the bounds from isocurvature fluctuations. If the axion is the DM particle, we have obtained its mass as a function of the initial abundance and mass of the PBH population allowed by the BBN and CMB bounds on the theoretical model, as shown in figure~\ref{fig:densityplot}. We have commented on the implication of the results in light of future searches for light axions in superconducting cavities. Other production channels such as PBH evaporation and thermal decoupling might lead to an additional dark radiation component that can show up in future CMB surveys.

This model leads to an effective early matter-dominated epoch; its effects on the DM axion have been rigorously discussed in literature~\cite{Visinelli:2009kt}. Overall, our new contributions can be summed up by two main findings. First, the results provided in figure~\ref{fig:densityplot} facilitate us to draw a relation between the PBH parameters and axion mass, and they serve as a guideline in models inclusive of inflation where both these entities are predicted. Second, the scenario is testable in the near future through indirect probes such as GW detectors~\cite{Seto:2001qf, Yagi:2011wg, Maggiore:2019uih, 2017arXiv170200786A}, CMB Stage IV experiments, and pulsar timing arrays, besides the direct search for light axions. This direct search with superconducting cavities are ongoing.~\cite{Berlin:2019ahk, Berlin:2020vrk, Berlin:2022hfx}.

Moreover, another merit of the model is it accommodates a scale of inflation that is higher than what is generally demanded by the cosmology of the QCD axion, as detailed in figure~\ref{fig:Plotmassvstheta}. The inflation scale $H_I \sim 10^{12}\,$GeV corroborates the model in cases where the axion mass is as light as $m_0 \sim 10^{-8}\,$eV.

\section*{Acknowledgements}

We thank Tanumoy Mandal and Nicol{\'a}s Bernal for their valuable insights and constructive comments throughout the completion of this work. LV thanks Chen Sun, Ying-Ying Li, and Ryan Janish for useful discussions on the detection of light bosons. We acknowledge the high-performance computing time at the Padmanabha cluster, IISER Thiruvananthapuram, India.

\appendix

\section{Numerical fit of the function $g_*(T)$}
\label{sec:dof}

In this work, we have derived the number of relativistic degrees of freedom $g_*(T)$ using the expressions from Ref.~\cite{Coleman:2003hs} for a Standard Model scenario augmented with an axion degree of freedom using the expression
\begin{equation}
    g_*(T) = \sum_i\left(\frac{T}{T_i}\right)^4\,g_i\frac{15}{\pi^4}\int_{x_i}^{+\infty}{\rm d}u\frac{u^2\sqrt{u^2-x_i^2}}{\exp(u)\pm 1}\,,
\end{equation}
where the sum runs over all species $i$ with mass $m_i$ and degrees of freedom $g_i$, with the sign $+$ ($-$) is intended for fermions (bosons) and with $x_i \equiv m_i/T$.\footnote{See Ref.~\cite{Husdal:2016haj} for the values of $g_i$ in the SM.} At temperatures well above the confinement scale $T_C$, we have taken the masses of the SM content from the 2022 release of the Particle Data Group~\cite{Workman:2022ynf} and we have neglected the masses of the neutrinos. Below the confinement scale, we account for the effect of all mesons and baryons of mass below $3\,$GeV. Setting $t = \ln(T/{\rm GeV})$, we fit our numerical result over the range $T \in [1{\rm\,MeV}, 1{\rm\,TeV}]$ to the function
\begin{equation}
    \label{eq:gstar}
    g_*(T) = a^{(0)} + \sum_{j=1}^5 a^{(1)}_j\left(1 + \tanh\frac{t - a^{(2)}_j}{a^{(3)}_j}\right)\,,
\end{equation}
which is inspired from Ref.~\cite{Wantz:2009it}. In table~\ref{tabfit} we report the results obtained.
\begin{table}
\centering
\def\arraystretch{1.5}
	\begin{tabular}{c|ccccc}
    \cline{2-5}
    \hline
    $j$ & 1 & 2 & 3 & 4 & 5\\
    \hline
    \hline
    $a^{(0)}$ & \multicolumn{5}{c}{39.77} \\
    $a^{(1)}_j$ &  33.99 & -10.40 & -20.52 & -6.81 & -10.26\\
    $a^{(2)}_j$ & -891.51 &  3.82 &  -1.72 & -2.35 &  -0.11\\
    $a^{(3)}_j$ & 437.30 &  -0.88 &  -0.15 & -1.22 &  -0.95\\
    \hline\hline
	\end{tabular}
	\caption{Coefficients obtained for the numerical fit of the function $g_*(T)$ in Eq.~\eqref{eq:gstar}.}
	\label{tabfit}
\end{table}

\section{Equations used to solve for the background cosmology}
\label{sec:numerics}

Here, we explicitly show the modifications in equations which describe the evolution of various components. We make these modifications to obtain a set of expressions, that is more suitable to be solved numerically.

\subsection{Cosmological background}
\label{sec:background}

We rewrite Eq.~\eqref{eq:rhovsT} in a form that is more suitable for numerical computations, using
\begin{equation}
    -\frac{{\rm d}M_{\rm BH}/{\rm d}t}{M_{\rm BH}} \equiv \frac{1}{t_{\rm evap}}\,.
\end{equation}
We introduce the quantities $x \equiv \ln (\rho_r/\bar\rho)$ and $y \equiv \ln (\rho_{\rm BH}/\bar\rho)$, so that the Hubble rate in terms of the scale factor $a$ becomes
\begin{equation}
    \label{eq:HubbleRescaled}
    H \equiv \frac{1}{a}\frac{{\rm d}a}{{\rm d}t} = \sqrt{\frac{8\pi G_N \bar\rho}{3}}(e^y + e^x)^{1/2}\,,
\end{equation}
and we fix the value of $\bar \rho$ so that $(8\pi G_N\bar\rho/3) t_{\rm evap}^2 = 1$. Temperature is rescaled according to its value at PBH formation $T_f$ in Eq.~\eqref{eq:Tinitial} by introducing $\tau \equiv \ln (T/T_f)$, so that the quantity $x \equiv \ln (\rho_r/\bar\rho)$ is
\begin{equation}
    x = 4\tau + \ln(\rho_r(T_f)/\bar\rho) + \ln(g_*(T)/g_*(T_f))\,.
\end{equation}
With these definitions, Eqs.~\eqref{eq:hubble} and~\eqref{eq:rhovsT} are written as
\begin{eqnarray}
    \left[(e^y \!+\! e^x)^{1/2} \!-\! \frac{e^{y-x}}{4}\right]Q(\tau)\frac{{\rm d} b}{{\rm d}\tau} &=& -(e^y \!+\! e^x)^{1/2}\!,\label{eq:hubble2}\\
    \left[(e^y \!+\! e^x)^{1/2} \!-\! \frac{e^{y-x}}{4}\right]Q(\tau)\frac{{\rm d} y}{{\rm d}\tau} &=& 1 \!+\! 3(e^y \!+\! e^x)^{1/2}\!\!, \label{eq:rhovsT2}
\end{eqnarray}
where $b \equiv \ln a$, and the expression for $Q(T)$ in Eq.~\eqref{def:QT} is
\begin{equation}
    \label{def:QT1}
    Q(\tau) \equiv \left(1 + \frac{1}{4g_*(T)}\frac{{\rm d} g_*(T)}{{\rm d}\tau}\right)^{-1}\,.
\end{equation}
The initial condition for $y$ is set at $T = T_f$ or $\tau = 0$, and it is obtained from Eq.~\eqref{eq:definebeta} as
\begin{equation}
    y(0) = \ln(\beta\rho_{\rm crit}(t_f)/\bar\rho) = \ln(\beta H_f^2t_{\rm evap}^2)\,,
\end{equation}
with $H_f = \gamma/(2G_N M_{\rm BH})$ and $\rho_{\rm crit}(t_f) = 3H_f^2/(8\pi G_N)$, and the initial condition for the scale factor is $b = 0$ at $\tau = 0$. These expressions allow for a speedy resolution for the content of the Universe well into PBH domination and through its evaporation for various choices of the PBH parameters, due to their logarithmic dependence on both temperature and energy densities.

\subsection{Equations for the axion field}
\label{sec:axionEOM}

The axion field is set in motion when the mass of the particle is of the same order as the Hubble rate. We first define the temperature $T_{\rm osc}$ around which the field oscillations begin and which is also the solution to equation $m_\phi(T_{\rm osc}) = H(T_{\rm osc})$. The value of $T_{\rm osc}$ obtained is used to rescale the independent variable and to facilitate the numerical resolution of the equation of motion. With rescaling the Hubble rate given in Eq.~\eqref{eq:HubbleRescaled} and solving Eqs.~\eqref{eq:hubble2}--\eqref{eq:rhovsT2}, we obtain an implicit equation which reads as
\begin{equation}
    (m_0 \, t_{\rm evap})^2\,\tilde\chi(T_{\rm osc}) = e^{y_{\rm osc}} \!+\! e^{x_{\rm osc}}\,,
\end{equation}
where $y_{\rm osc}$ and $x_{\rm osc}$ are the values of the quantities $y$ and $x$ evaluated at $T_{\rm osc}$. The expression leads to a value $T_{\rm osc}$ which is a function of the axion mass at zero temperature $m_0$ and the initial axion angle $\theta_i$, other than the PBH parameters.

Since both the axion mass and the function $g_*(T)$ depend on temperature, it is far better, neat, and useful to express all quantities as functions of $T$. The independent variable $T$ has further been rescaled to take $T_{\rm osc}$ into account. For this, we introduce the independent variable $\xi \equiv \ln(T/T_{\rm osc})$, related to the variable $\tau$ by $\tau \equiv \ln(T/T_f) = \xi + \ln(T_{\rm osc}/T_f)$. We express the yield in Eq.~\eqref{eq:Y} in terms of $\xi$ by solving. The axion oscillations have also begun.
\begin{equation}
    \label{eq:Y1}
    (\ln Y_\phi)' = -\frac{3}{4}\,\frac{g_*(T_{\rm osc})}{g_S(T)}\,b'\,\frac{\bar\rho}{\rho_r(T_{\rm osc})}\,\frac{e^{y - 4 \xi}}{\sqrt{e^y + e^x}}\,,
\end{equation}
where a prime is a differential with respect to $\xi$. The solution of Eq.~\eqref{eq:Y1} leaves the initial value of $Y_\phi$ undetermined, which is why we need to solve the equation of motion for a DM axion. 

The coherent oscillations in the QCD axion field are described by Eq.~\eqref{eq:motion} which, for superhorizon modes with the axion angle as $\theta \equiv \phi / f_{\rm PQ}$, reads
\begin{equation}
    \label{eq:motionA}
    \ddot \theta + 3H\dot\theta + \frac{1}{f_{\rm PQ}^2}\frac{\partial V(\theta)}{\partial \theta} = 0\,,
\end{equation}
with the axion potential given in Eq.~\eqref{eq:axionpotential}. Once Eq.~\eqref{eq:motionA} is expressed in terms of the variable $\xi$ we obtain
\begin{equation}
    \theta'' + \left(\frac{H'}{H} + 3 b'-\frac{b''}{b'}\right) \theta' + \left(\frac{b'}{H}\right)^2\frac{1}{f_{\rm PQ}^2}\frac{\partial V}{\partial \theta} = 0\,.
\end{equation}
Note, that since the scale factor enters only through the derivatives of the function $b \equiv \ln a$, a shift in the normalization of the scale factor is irrelevant to the equations of motion. Since $V(\theta)$ is given in Eq.~\eqref{eq:Tdependentmass} and the Hubble rate is expressed in Eq.~\eqref{eq:HubbleRescaled}, we finally obtain
\begin{equation}
    \theta'' \!+\! \left(\frac{H'}{H} \!+\! 3 b' \!-\! \frac{b''}{b'}\right) \theta' \!+ \frac{e^{y_{\rm osc}} \!+\! e^{x_{\rm osc}}}{e^y \!+\! e^x}\frac{\tilde\chi(T)}{\tilde\chi(T_{\rm osc})}(b')^2\!\sin\theta = 0\,.
\end{equation}
The expression above depends on the axion mass $m_0$ and the PBH mass through $t_{\rm evap}$. The idea of rescaling the equation of motion for the axion field around $T_{\rm osc}$ is certainly not new, see Eq.~(3.5) in Ref.~\cite{Kolb:1993hw}, however here we give it a new interpretation related to the background cosmology which contains the decaying component from PBH evaporation. We solve the expression above with the initial condition $\theta = \theta_i$ for $T \gg T_{\rm osc}$. The energy density of the axion field in Eq.~\eqref{eq:rhophi} is rewritten as
\begin{eqnarray}
    \rho_\phi &=& \frac{\Lambda^4}{m_0^2}\left[\frac{1}{2}H^2\left(\frac{\theta'}{b'}\right)^2 + m_\phi^2(T)(1-\cos\theta)\right]\\
    &=& \Lambda^4\!\left[\frac{e^y + e^x}{2(m_0 \, t_{\rm evap})^2}\left(\frac{\theta'}{b'}\right)^2 \!+\! \tilde\chi(T)(1 \!-\! \cos\theta)\right],\nonumber
\end{eqnarray}
which allows the computation of the present DM density, visit Eq.~\eqref{eq:axionDM} for more.

\bibliographystyle{JHEP.bst}
\bibliography{AxionPBH.bib}

\providecommand{\href}[2]{#2}\begingroup\raggedright\begin{thebibliography}{100}

\bibitem{Ghez:1998ph}
A.M.~Ghez, B.L.~Klein, M.~Morris and E.E.~Becklin, \emph{{High proper motion
  stars in the vicinity of Sgr A*: Evidence for a supermassive black hole at
  the center of our galaxy}},
  \href{https://doi.org/10.1086/306528}{\emph{Astrophys. J.} {\bfseries 509}
  (1998) 678} [\href{https://arxiv.org/abs/astro-ph/9807210}{{\ttfamily
  astro-ph/9807210}}].

\bibitem{Ghez:2008ms}
A.M.~Ghez et~al., \emph{{Measuring Distance and Properties of the Milky Way's
  Central Supermassive Black Hole with Stellar Orbits}},
  \href{https://doi.org/10.1086/592738}{\emph{Astrophys. J.} {\bfseries 689}
  (2008) 1044} [\href{https://arxiv.org/abs/0808.2870}{{\ttfamily 0808.2870}}].

\bibitem{Gillessen:2008qv}
S.~Gillessen, F.~Eisenhauer, S.~Trippe, T.~Alexander, R.~Genzel, F.~Martins
  et~al., \emph{{Monitoring stellar orbits around the Massive Black Hole in the
  Galactic Center}},
  \href{https://doi.org/10.1088/0004-637X/692/2/1075}{\emph{Astrophys. J.}
  {\bfseries 692} (2009) 1075}
  [\href{https://arxiv.org/abs/0810.4674}{{\ttfamily 0810.4674}}].

\bibitem{LIGOScientific:2016aoc}
{\scshape LIGO Scientific, Virgo} collaboration, \emph{{Observation of
  Gravitational Waves from a Binary Black Hole Merger}},
  \href{https://doi.org/10.1103/PhysRevLett.116.061102}{\emph{Phys. Rev. Lett.}
  {\bfseries 116} (2016) 061102}
  [\href{https://arxiv.org/abs/1602.03837}{{\ttfamily 1602.03837}}].

\bibitem{LIGOScientific:2018mvr}
{\scshape LIGO Scientific, Virgo} collaboration, \emph{{GWTC-1: A
  Gravitational-Wave Transient Catalog of Compact Binary Mergers Observed by
  LIGO and Virgo during the First and Second Observing Runs}},
  \href{https://doi.org/10.1103/PhysRevX.9.031040}{\emph{Phys. Rev. X}
  {\bfseries 9} (2019) 031040}
  [\href{https://arxiv.org/abs/1811.12907}{{\ttfamily 1811.12907}}].

\bibitem{EventHorizonTelescope:2019dse}
{\scshape Event Horizon Telescope} collaboration, \emph{{First M87 Event
  Horizon Telescope Results. I. The Shadow of the Supermassive Black Hole}},
  \href{https://doi.org/10.3847/2041-8213/ab0ec7}{\emph{Astrophys. J. Lett.}
  {\bfseries 875} (2019) L1}
  [\href{https://arxiv.org/abs/1906.11238}{{\ttfamily 1906.11238}}].

\bibitem{EventHorizonTelescope:2022xnr}
{\scshape Event Horizon Telescope} collaboration, \emph{{First Sagittarius A*
  Event Horizon Telescope Results. I. The Shadow of the Supermassive Black Hole
  in the Center of the Milky Way}},
  \href{https://doi.org/10.3847/2041-8213/ac6674}{\emph{Astrophys. J. Lett.}
  {\bfseries 930} (2022) L12}.

\bibitem{Heger:2002by}
A.~Heger, C.L.~Fryer, S.E.~Woosley, N.~Langer and D.H.~Hartmann, \emph{{How
  massive single stars end their life}},
  \href{https://doi.org/10.1086/375341}{\emph{Astrophys. J.} {\bfseries 591}
  (2003) 288} [\href{https://arxiv.org/abs/astro-ph/0212469}{{\ttfamily
  astro-ph/0212469}}].

\bibitem{Carr:1974nx}
B.J.~Carr and S.W.~Hawking, \emph{{Black holes in the early Universe}},
  {\emph{Mon. Not. Roy. Astron. Soc.} {\bfseries 168} (1974) 399}.

\bibitem{Carr:1975qj}
B.J.~Carr, \emph{{The Primordial black hole mass spectrum}},
  \href{https://doi.org/10.1086/153853}{\emph{Astrophys. J.} {\bfseries 201}
  (1975) 1}.

\bibitem{Zeldovich:1967lct}
Y.B.~Zel'dovich and I.D.~Novikov, \emph{{The Hypothesis of Cores Retarded
  during Expansion and the Hot Cosmological Model}}, {\emph{Soviet Astron. AJ
  (Engl. Transl. ),} {\bfseries 10} (1967) 602}.

\bibitem{Shibata:1999zs}
M.~Shibata and M.~Sasaki, \emph{{Black hole formation in the Friedmann
  universe: Formulation and computation in numerical relativity}},
  \href{https://doi.org/10.1103/PhysRevD.60.084002}{\emph{Phys. Rev. D}
  {\bfseries 60} (1999) 084002}
  [\href{https://arxiv.org/abs/gr-qc/9905064}{{\ttfamily gr-qc/9905064}}].

\bibitem{Niemeyer:1999ak}
J.C.~Niemeyer and K.~Jedamzik, \emph{{Dynamics of primordial black hole
  formation}}, \href{https://doi.org/10.1103/PhysRevD.59.124013}{\emph{Phys.
  Rev. D} {\bfseries 59} (1999) 124013}
  [\href{https://arxiv.org/abs/astro-ph/9901292}{{\ttfamily
  astro-ph/9901292}}].

\bibitem{Musco:2004ak}
I.~Musco, J.C.~Miller and L.~Rezzolla, \emph{{Computations of primordial black
  hole formation}},
  \href{https://doi.org/10.1088/0264-9381/22/7/013}{\emph{Class. Quant. Grav.}
  {\bfseries 22} (2005) 1405}
  [\href{https://arxiv.org/abs/gr-qc/0412063}{{\ttfamily gr-qc/0412063}}].

\bibitem{Wu:2020ilx}
Y.-P.~Wu, \emph{{Peak statistics for the primordial black hole abundance}},
  \href{https://doi.org/10.1016/j.dark.2020.100654}{\emph{Phys. Dark Univ.}
  {\bfseries 30} (2020) 100654}
  [\href{https://arxiv.org/abs/2005.00441}{{\ttfamily 2005.00441}}].

\bibitem{Gow:2020bzo}
A.D.~Gow, C.T.~Byrnes, P.S.~Cole and S.~Young, \emph{{The power spectrum on
  small scales: Robust constraints and comparing PBH methodologies}},
  \href{https://doi.org/10.1088/1475-7516/2021/02/002}{\emph{JCAP} {\bfseries
  02} (2021) 002} [\href{https://arxiv.org/abs/2008.03289}{{\ttfamily
  2008.03289}}].

\bibitem{Garcia-Bellido:1996mdl}
J.~Garcia-Bellido, A.D.~Linde and D.~Wands, \emph{{Density perturbations and
  black hole formation in hybrid inflation}},
  \href{https://doi.org/10.1103/PhysRevD.54.6040}{\emph{Phys. Rev. D}
  {\bfseries 54} (1996) 6040}
  [\href{https://arxiv.org/abs/astro-ph/9605094}{{\ttfamily
  astro-ph/9605094}}].

\bibitem{Flores:2020drq}
M.M.~Flores and A.~Kusenko, \emph{{Primordial Black Holes from Long-Range
  Scalar Forces and Scalar Radiative Cooling}},
  \href{https://doi.org/10.1103/PhysRevLett.126.041101}{\emph{Phys. Rev. Lett.}
  {\bfseries 126} (2021) 041101}
  [\href{https://arxiv.org/abs/2008.12456}{{\ttfamily 2008.12456}}].

\bibitem{Maeso:2021xvl}
D.N.~Maeso, L.~Marzola, M.~Raidal, V.~Vaskonen and H.~Veerm\"ae,
  \emph{{Primordial black holes from spectator field bubbles}},
  \href{https://doi.org/10.1088/1475-7516/2022/02/017}{\emph{JCAP} {\bfseries
  02} (2022) 017} [\href{https://arxiv.org/abs/2112.01505}{{\ttfamily
  2112.01505}}].

\bibitem{Byrnes:2018clq}
C.T.~Byrnes, M.~Hindmarsh, S.~Young and M.R.S.~Hawkins, \emph{{Primordial black
  holes with an accurate QCD equation of state}},
  \href{https://doi.org/10.1088/1475-7516/2018/08/041}{\emph{JCAP} {\bfseries
  08} (2018) 041} [\href{https://arxiv.org/abs/1801.06138}{{\ttfamily
  1801.06138}}].

\bibitem{Carr:2020gox}
B.~Carr, K.~Kohri, Y.~Sendouda and J.~Yokoyama, \emph{{Constraints on
  primordial black holes}},
  \href{https://doi.org/10.1088/1361-6633/ac1e31}{\emph{Rept. Prog. Phys.}
  {\bfseries 84} (2021) 116902}
  [\href{https://arxiv.org/abs/2002.12778}{{\ttfamily 2002.12778}}].

\bibitem{Carr:2020xqk}
B.~Carr and F.~Kuhnel, \emph{{Primordial Black Holes as Dark Matter: Recent
  Developments}},
  \href{https://doi.org/10.1146/annurev-nucl-050520-125911}{\emph{Ann. Rev.
  Nucl. Part. Sci.} {\bfseries 70} (2020) 355}
  [\href{https://arxiv.org/abs/2006.02838}{{\ttfamily 2006.02838}}].

\bibitem{Hawking:1974rv}
S.W.~Hawking, \emph{{Black hole explosions}},
  \href{https://doi.org/10.1038/248030a0}{\emph{Nature} {\bfseries 248} (1974)
  30}.

\bibitem{Ali-Haimoud:2016mbv}
Y.~Ali-Ha\"\i{}moud and M.~Kamionkowski, \emph{{Cosmic microwave background
  limits on accreting primordial black holes}},
  \href{https://doi.org/10.1103/PhysRevD.95.043534}{\emph{Phys. Rev. D}
  {\bfseries 95} (2017) 043534}
  [\href{https://arxiv.org/abs/1612.05644}{{\ttfamily 1612.05644}}].

\bibitem{Acharya:2020jbv}
S.K.~Acharya and R.~Khatri, \emph{{CMB and BBN constraints on evaporating
  primordial black holes revisited}},
  \href{https://doi.org/10.1088/1475-7516/2020/06/018}{\emph{JCAP} {\bfseries
  06} (2020) 018} [\href{https://arxiv.org/abs/2002.00898}{{\ttfamily
  2002.00898}}].

\bibitem{Niikura:2017zjd}
H.~Niikura et~al., \emph{{Microlensing constraints on primordial black holes
  with Subaru/HSC Andromeda observations}},
  \href{https://doi.org/10.1038/s41550-019-0723-1}{\emph{Nature Astron.}
  {\bfseries 3} (2019) 524} [\href{https://arxiv.org/abs/1701.02151}{{\ttfamily
  1701.02151}}].

\bibitem{Carr:1997cn}
B.J.~Carr and M.~Sakellariadou, \emph{{Dynamical constraints on dark compact
  objects}}, \href{https://doi.org/10.1086/307071}{\emph{Astrophys. J.}
  {\bfseries 516} (1999) 195}.

\bibitem{Carr:2020erq}
B.~Carr, F.~Kuhnel and L.~Visinelli, \emph{{Constraints on Stupendously Large
  Black Holes}}, \href{https://doi.org/10.1093/mnras/staa3651}{\emph{Mon. Not.
  Roy. Astron. Soc.} {\bfseries 501} (2021) 2029}
  [\href{https://arxiv.org/abs/2008.08077}{{\ttfamily 2008.08077}}].

\bibitem{Coogan:2020tuf}
A.~Coogan, L.~Morrison and S.~Profumo, \emph{{Direct Detection of Hawking
  Radiation from Asteroid-Mass Primordial Black Holes}},
  \href{https://doi.org/10.1103/PhysRevLett.126.171101}{\emph{Phys. Rev. Lett.}
  {\bfseries 126} (2021) 171101}
  [\href{https://arxiv.org/abs/2010.04797}{{\ttfamily 2010.04797}}].

\bibitem{Carr:2016drx}
B.~Carr, F.~Kuhnel and M.~Sandstad, \emph{{Primordial Black Holes as Dark
  Matter}}, \href{https://doi.org/10.1103/PhysRevD.94.083504}{\emph{Phys. Rev.
  D} {\bfseries 94} (2016) 083504}
  [\href{https://arxiv.org/abs/1607.06077}{{\ttfamily 1607.06077}}].

\bibitem{Green:2020jor}
A.M.~Green and B.J.~Kavanagh, \emph{{Primordial Black Holes as a dark matter
  candidate}}, \href{https://doi.org/10.1088/1361-6471/abc534}{\emph{J. Phys.
  G} {\bfseries 48} (2021) 043001}
  [\href{https://arxiv.org/abs/2007.10722}{{\ttfamily 2007.10722}}].

\bibitem{Chakraborty:2022mwu}
A.~Chakraborty, P.K.~Chanda, K.L.~Pandey and S.~Das, \emph{{Formation and
  Abundance of Late-forming Primordial Black Holes as Dark Matter}},
  \href{https://doi.org/10.3847/1538-4357/ac6ddd}{\emph{Astrophys. J.}
  {\bfseries 932} (2022) 119}
  [\href{https://arxiv.org/abs/2204.09628}{{\ttfamily 2204.09628}}].

\bibitem{Toussaint:1978br}
D.~Toussaint, S.B.~Treiman, F.~Wilczek and A.~Zee, \emph{{Matter - Antimatter
  Accounting, Thermodynamics, and Black Hole Radiation}},
  \href{https://doi.org/10.1103/PhysRevD.19.1036}{\emph{Phys. Rev. D}
  {\bfseries 19} (1979) 1036}.

\bibitem{Turner:1979zj}
M.S.~Turner and D.N.~Schramm, \emph{{The Origin of Baryons in the Universe and
  the Astrophysical Implications}},
  \href{https://doi.org/10.1038/279303a0}{\emph{Nature} {\bfseries 279} (1979)
  303}.

\bibitem{Nesseris:2019fwr}
S.~Nesseris, D.~Sapone and S.~Sypsas, \emph{{Evaporating primordial black holes
  as varying dark energy}},
  \href{https://doi.org/10.1016/j.dark.2019.100413}{\emph{Phys. Dark Univ.}
  {\bfseries 27} (2020) 100413}
  [\href{https://arxiv.org/abs/1907.05608}{{\ttfamily 1907.05608}}].

\bibitem{DiValentino:2021izs}
E.~Di~Valentino, O.~Mena, S.~Pan, L.~Visinelli, W.~Yang, A.~Melchiorri et~al.,
  \emph{{In the realm of the Hubble tension\textemdash{}a review of
  solutions}}, \href{https://doi.org/10.1088/1361-6382/ac086d}{\emph{Class.
  Quant. Grav.} {\bfseries 38} (2021) 153001}
  [\href{https://arxiv.org/abs/2103.01183}{{\ttfamily 2103.01183}}].

\bibitem{Jungman:1995df}
G.~Jungman, M.~Kamionkowski and K.~Griest, \emph{{Supersymmetric dark matter}},
  \href{https://doi.org/10.1016/0370-1573(95)00058-5}{\emph{Phys. Rept.}
  {\bfseries 267} (1996) 195}
  [\href{https://arxiv.org/abs/hep-ph/9506380}{{\ttfamily hep-ph/9506380}}].

\bibitem{Fujita:2014hha}
T.~Fujita, M.~Kawasaki, K.~Harigaya and R.~Matsuda, \emph{{Baryon asymmetry,
  dark matter, and density perturbation from primordial black holes}},
  \href{https://doi.org/10.1103/PhysRevD.89.103501}{\emph{Phys. Rev. D}
  {\bfseries 89} (2014) 103501}
  [\href{https://arxiv.org/abs/1401.1909}{{\ttfamily 1401.1909}}].

\bibitem{Allahverdi:2017sks}
R.~Allahverdi, J.~Dent and J.~Osinski, \emph{{Nonthermal production of dark
  matter from primordial black holes}},
  \href{https://doi.org/10.1103/PhysRevD.97.055013}{\emph{Phys. Rev. D}
  {\bfseries 97} (2018) 055013}
  [\href{https://arxiv.org/abs/1711.10511}{{\ttfamily 1711.10511}}].

\bibitem{Lennon:2017tqq}
O.~Lennon, J.~March-Russell, R.~Petrossian-Byrne and H.~Tillim, \emph{{Black
  Hole Genesis of Dark Matter}},
  \href{https://doi.org/10.1088/1475-7516/2018/04/009}{\emph{JCAP} {\bfseries
  04} (2018) 009} [\href{https://arxiv.org/abs/1712.07664}{{\ttfamily
  1712.07664}}].

\bibitem{Morrison:2018xla}
L.~Morrison, S.~Profumo and Y.~Yu, \emph{{Melanopogenesis: Dark Matter of
  (almost) any Mass and Baryonic Matter from the Evaporation of Primordial
  Black Holes weighing a Ton (or less)}},
  \href{https://doi.org/10.1088/1475-7516/2019/05/005}{\emph{JCAP} {\bfseries
  05} (2019) 005} [\href{https://arxiv.org/abs/1812.10606}{{\ttfamily
  1812.10606}}].

\bibitem{Hooper:2019gtx}
D.~Hooper, G.~Krnjaic and S.D.~McDermott, \emph{{Dark Radiation and Superheavy
  Dark Matter from Black Hole Domination}},
  \href{https://doi.org/10.1007/JHEP08(2019)001}{\emph{JHEP} {\bfseries 08}
  (2019) 001} [\href{https://arxiv.org/abs/1905.01301}{{\ttfamily
  1905.01301}}].

\bibitem{Masina:2020xhk}
I.~Masina, \emph{{Dark matter and dark radiation from evaporating primordial
  black holes}},
  \href{https://doi.org/10.1140/epjp/s13360-020-00564-9}{\emph{Eur. Phys. J.
  Plus} {\bfseries 135} (2020) 552}
  [\href{https://arxiv.org/abs/2004.04740}{{\ttfamily 2004.04740}}].

\bibitem{Cheek:2021cfe}
A.~Cheek, L.~Heurtier, Y.F.~Perez-Gonzalez and J.~Turner, \emph{{Primordial
  black hole evaporation and dark matter production. II. Interplay with the
  freeze-in or freeze-out mechanism}},
  \href{https://doi.org/10.1103/PhysRevD.105.015023}{\emph{Phys. Rev. D}
  {\bfseries 105} (2022) 015023}
  [\href{https://arxiv.org/abs/2107.00016}{{\ttfamily 2107.00016}}].

\bibitem{Chattopadhyay:2022fwa}
P.~Chattopadhyay, A.~Chaudhuri and M.Y.~Khlopov, \emph{{Dark Matter from
  Evaporating PBH dominated in the Early Universe}},
  \href{https://arxiv.org/abs/2209.11288}{{\ttfamily 2209.11288}}.

\bibitem{Baldes:2020nuv}
I.~Baldes, Q.~Decant, D.C.~Hooper and L.~Lopez-Honorez, \emph{{Non-Cold Dark
  Matter from Primordial Black Hole Evaporation}},
  \href{https://doi.org/10.1088/1475-7516/2020/08/045}{\emph{JCAP} {\bfseries
  08} (2020) 045} [\href{https://arxiv.org/abs/2004.14773}{{\ttfamily
  2004.14773}}].

\bibitem{Gondolo:2020uqv}
P.~Gondolo, P.~Sandick and B.~Shams Es~Haghi, \emph{{Effects of primordial
  black holes on dark matter models}},
  \href{https://doi.org/10.1103/PhysRevD.102.095018}{\emph{Phys. Rev. D}
  {\bfseries 102} (2020) 095018}
  [\href{https://arxiv.org/abs/2009.02424}{{\ttfamily 2009.02424}}].

\bibitem{Boucenna:2017ghj}
S.M.~Boucenna, F.~Kuhnel, T.~Ohlsson and L.~Visinelli, \emph{{Novel Constraints
  on Mixed Dark-Matter Scenarios of Primordial Black Holes and WIMPs}},
  \href{https://doi.org/10.1088/1475-7516/2018/07/003}{\emph{JCAP} {\bfseries
  07} (2018) 003} [\href{https://arxiv.org/abs/1712.06383}{{\ttfamily
  1712.06383}}].

\bibitem{Bertone:2019vsk}
G.~Bertone, A.M.~Coogan, D.~Gaggero, B.J.~Kavanagh and C.~Weniger,
  \emph{{Primordial Black Holes as Silver Bullets for New Physics at the Weak
  Scale}}, \href{https://doi.org/10.1103/PhysRevD.100.123013}{\emph{Phys. Rev.
  D} {\bfseries 100} (2019) 123013}
  [\href{https://arxiv.org/abs/1905.01238}{{\ttfamily 1905.01238}}].

\bibitem{Hertzberg:2020hsz}
M.P.~Hertzberg, E.D.~Schiappacasse and T.T.~Yanagida, \emph{{Axion Star
  Nucleation in Dark Minihalos around Primordial Black Holes}},
  \href{https://doi.org/10.1103/PhysRevD.102.023013}{\emph{Phys. Rev. D}
  {\bfseries 102} (2020) 023013}
  [\href{https://arxiv.org/abs/2001.07476}{{\ttfamily 2001.07476}}].

\bibitem{Carr:2020mqm}
B.~Carr, F.~Kuhnel and L.~Visinelli, \emph{{Black holes and WIMPs: all or
  nothing or something else}},
  \href{https://doi.org/10.1093/mnras/stab1930}{\emph{Mon. Not. Roy. Astron.
  Soc.} {\bfseries 506} (2021) 3648}
  [\href{https://arxiv.org/abs/2011.01930}{{\ttfamily 2011.01930}}].

\bibitem{Gines:2022qzy}
E.U.~Gin\'es, O.~Mena and S.J.~Witte, \emph{{Revisiting constraints on WIMPs
  around primordial black holes}},
  \href{https://doi.org/10.1103/PhysRevD.106.063538}{\emph{Phys. Rev. D}
  {\bfseries 106} (2022) 063538}
  [\href{https://arxiv.org/abs/2207.09481}{{\ttfamily 2207.09481}}].

\bibitem{Chanda:2022hls}
P.~Chanda, J.~Scholtz and J.~Unwin, \emph{{Improved Constraints on Dark Matter
  Annihilations Around Primordial Black Holes}},
  \href{https://arxiv.org/abs/2209.07541}{{\ttfamily 2209.07541}}.

\bibitem{Chung:1998rq}
D.J.H.~Chung, E.W.~Kolb and A.~Riotto, \emph{{Production of massive particles
  during reheating}},
  \href{https://doi.org/10.1103/PhysRevD.60.063504}{\emph{Phys. Rev. D}
  {\bfseries 60} (1999) 063504}
  [\href{https://arxiv.org/abs/hep-ph/9809453}{{\ttfamily hep-ph/9809453}}].

\bibitem{Moroi:1999zb}
T.~Moroi and L.~Randall, \emph{{Wino cold dark matter from anomaly mediated
  SUSY breaking}},
  \href{https://doi.org/10.1016/S0550-3213(99)00748-8}{\emph{Nucl. Phys. B}
  {\bfseries 570} (2000) 455}
  [\href{https://arxiv.org/abs/hep-ph/9906527}{{\ttfamily hep-ph/9906527}}].

\bibitem{Gelmini:2008sh}
G.B.~Gelmini and P.~Gondolo, \emph{{Ultra-cold WIMPs: relics of non-standard
  pre-BBN cosmologies}},
  \href{https://doi.org/10.1088/1475-7516/2008/10/002}{\emph{JCAP} {\bfseries
  10} (2008) 002} [\href{https://arxiv.org/abs/0803.2349}{{\ttfamily
  0803.2349}}].

\bibitem{Visinelli:2015eka}
L.~Visinelli and P.~Gondolo, \emph{{Kinetic decoupling of WIMPs: analytic
  expressions}}, \href{https://doi.org/10.1103/PhysRevD.91.083526}{\emph{Phys.
  Rev. D} {\bfseries 91} (2015) 083526}
  [\href{https://arxiv.org/abs/1501.02233}{{\ttfamily 1501.02233}}].

\bibitem{Visinelli:2017qga}
L.~Visinelli, \emph{{(Non-)thermal production of WIMPs during kination}},
  \href{https://doi.org/10.3390/sym10110546}{\emph{Symmetry} {\bfseries 10}
  (2018) 546} [\href{https://arxiv.org/abs/1710.11006}{{\ttfamily
  1710.11006}}].

\bibitem{Weinberg:1977ma}
S.~Weinberg, \emph{{A New Light Boson?}},
  \href{https://doi.org/10.1103/PhysRevLett.40.223}{\emph{Phys. Rev. Lett.}
  {\bfseries 40} (1978) 223}.

\bibitem{Wilczek:1977pj}
F.~Wilczek, \emph{{Problem of Strong $P$ and $T$ Invariance in the Presence of
  Instantons}}, \href{https://doi.org/10.1103/PhysRevLett.40.279}{\emph{Phys.
  Rev. Lett.} {\bfseries 40} (1978) 279}.

\bibitem{Peccei:1977ur}
R.D.~Peccei and H.R.~Quinn, \emph{{Constraints Imposed by CP Conservation in
  the Presence of Instantons}},
  \href{https://doi.org/10.1103/PhysRevD.16.1791}{\emph{Phys. Rev. D}
  {\bfseries 16} (1977) 1791}.

\bibitem{Peccei:1977hh}
R.D.~Peccei and H.R.~Quinn, \emph{{CP Conservation in the Presence of
  Instantons}}, \href{https://doi.org/10.1103/PhysRevLett.38.1440}{\emph{Phys.
  Rev. Lett.} {\bfseries 38} (1977) 1440}.

\bibitem{Preskill:1982cy}
J.~Preskill, M.B.~Wise and F.~Wilczek, \emph{{Cosmology of the Invisible
  Axion}}, \href{https://doi.org/10.1016/0370-2693(83)90637-8}{\emph{Phys.
  Lett. B} {\bfseries 120} (1983) 127}.

\bibitem{Abbott:1982af}
L.F.~Abbott and P.~Sikivie, \emph{{A Cosmological Bound on the Invisible
  Axion}}, \href{https://doi.org/10.1016/0370-2693(83)90638-X}{\emph{Phys.
  Lett. B} {\bfseries 120} (1983) 133}.

\bibitem{Dine:1982ah}
M.~Dine and W.~Fischler, \emph{{The Not So Harmless Axion}},
  \href{https://doi.org/10.1016/0370-2693(83)90639-1}{\emph{Phys. Lett. B}
  {\bfseries 120} (1983) 137}.

\bibitem{Lazarides:1987zf}
G.~Lazarides, C.~Panagiotakopoulos and Q.~Shafi, \emph{{Relaxing the
  Cosmological Bound on Axions}},
  \href{https://doi.org/10.1016/0370-2693(87)90115-8}{\emph{Phys. Lett. B}
  {\bfseries 192} (1987) 323}.

\bibitem{Lazarides:1990xp}
G.~Lazarides, R.K.~Schaefer, D.~Seckel and Q.~Shafi, \emph{{Dilution of
  Cosmological Axions by Entropy Production}},
  \href{https://doi.org/10.1016/0550-3213(90)90244-8}{\emph{Nucl. Phys. B}
  {\bfseries 346} (1990) 193}.

\bibitem{Visinelli:2009kt}
L.~Visinelli and P.~Gondolo, \emph{{Axion cold dark matter in non-standard
  cosmologies}}, \href{https://doi.org/10.1103/PhysRevD.81.063508}{\emph{Phys.
  Rev. D} {\bfseries 81} (2010) 063508}
  [\href{https://arxiv.org/abs/0912.0015}{{\ttfamily 0912.0015}}].

\bibitem{Visinelli:2017imh}
L.~Visinelli, \emph{{Light axion-like dark matter must be present during
  inflation}}, \href{https://doi.org/10.1103/PhysRevD.96.023013}{\emph{Phys.
  Rev. D} {\bfseries 96} (2017) 023013}
  [\href{https://arxiv.org/abs/1703.08798}{{\ttfamily 1703.08798}}].

\bibitem{Visinelli:2018wza}
L.~Visinelli and J.~Redondo, \emph{{Axion Miniclusters in Modified Cosmological
  Histories}}, \href{https://doi.org/10.1103/PhysRevD.101.023008}{\emph{Phys.
  Rev. D} {\bfseries 101} (2020) 023008}
  [\href{https://arxiv.org/abs/1808.01879}{{\ttfamily 1808.01879}}].

\bibitem{Nelson:2018via}
A.E.~Nelson and H.~Xiao, \emph{{Axion Cosmology with Early Matter Domination}},
  \href{https://doi.org/10.1103/PhysRevD.98.063516}{\emph{Phys. Rev. D}
  {\bfseries 98} (2018) 063516}
  [\href{https://arxiv.org/abs/1807.07176}{{\ttfamily 1807.07176}}].

\bibitem{Ramberg:2019dgi}
N.~Ramberg and L.~Visinelli, \emph{{Probing the Early Universe with Axion
  Physics and Gravitational Waves}},
  \href{https://doi.org/10.1103/PhysRevD.99.123513}{\emph{Phys. Rev. D}
  {\bfseries 99} (2019) 123513}
  [\href{https://arxiv.org/abs/1904.05707}{{\ttfamily 1904.05707}}].

\bibitem{Ramberg:2020oct}
N.~Ramberg and L.~Visinelli, \emph{{QCD axion and gravitational waves in light
  of NANOGrav results}},
  \href{https://doi.org/10.1103/PhysRevD.103.063031}{\emph{Phys. Rev. D}
  {\bfseries 103} (2021) 063031}
  [\href{https://arxiv.org/abs/2012.06882}{{\ttfamily 2012.06882}}].

\bibitem{Arias:2021rer}
P.~Arias, N.~Bernal, D.~Karamitros, C.~Maldonado, L.~Roszkowski and M.~Venegas,
  \emph{{New opportunities for axion dark matter searches in nonstandard
  cosmological models}},
  \href{https://doi.org/10.1088/1475-7516/2021/11/003}{\emph{JCAP} {\bfseries
  11} (2021) 003} [\href{https://arxiv.org/abs/2107.13588}{{\ttfamily
  2107.13588}}].

\bibitem{Bao:2022hsg}
Y.~Bao, J.~Fan and L.~Li, \emph{{Opening up window of post-inflationary QCD
  axion}},  \href{https://arxiv.org/abs/2209.09908}{{\ttfamily 2209.09908}}.

\bibitem{Arias:2022qjt}
P.~Arias, N.~Bernal, J.K.~Osi\'nski and L.~Roszkowski, \emph{{Dark Matter
  Axions in the Early Universe with a Period of Increasing Temperature}},
  \href{https://arxiv.org/abs/2207.07677}{{\ttfamily 2207.07677}}.

\bibitem{Bernal:2021yyb}
N.~Bernal, F.~Hajkarim and Y.~Xu, \emph{{Axion Dark Matter in the Time of
  Primordial Black Holes}},
  \href{https://doi.org/10.1103/PhysRevD.104.075007}{\emph{Phys. Rev. D}
  {\bfseries 104} (2021) 075007}
  [\href{https://arxiv.org/abs/2107.13575}{{\ttfamily 2107.13575}}].

\bibitem{Bernal:2021bbv}
N.~Bernal, Y.F.~Perez-Gonzalez, Y.~Xu and O.~Zapata, \emph{{ALP dark matter in
  a primordial black hole dominated universe}},
  \href{https://doi.org/10.1103/PhysRevD.104.123536}{\emph{Phys. Rev. D}
  {\bfseries 104} (2021) 123536}
  [\href{https://arxiv.org/abs/2110.04312}{{\ttfamily 2110.04312}}].

\bibitem{Mirbabayi:2019uph}
M.~Mirbabayi, A.~Gruzinov and J.~Nore\~na, \emph{{Spin of Primordial Black
  Holes}}, \href{https://doi.org/10.1088/1475-7516/2020/03/017}{\emph{JCAP}
  {\bfseries 03} (2020) 017}
  [\href{https://arxiv.org/abs/1901.05963}{{\ttfamily 1901.05963}}].

\bibitem{DeLuca:2019buf}
V.~De~Luca, V.~Desjacques, G.~Franciolini, A.~Malhotra and A.~Riotto,
  \emph{{The initial spin probability distribution of primordial black holes}},
  \href{https://doi.org/10.1088/1475-7516/2019/05/018}{\emph{JCAP} {\bfseries
  05} (2019) 018} [\href{https://arxiv.org/abs/1903.01179}{{\ttfamily
  1903.01179}}].

\bibitem{Dalianis:2018ymb}
I.~Dalianis, \emph{{Constraints on the curvature power spectrum from primordial
  black hole evaporation}},
  \href{https://doi.org/10.1088/1475-7516/2019/08/032}{\emph{JCAP} {\bfseries
  08} (2019) 032} [\href{https://arxiv.org/abs/1812.09807}{{\ttfamily
  1812.09807}}].

\bibitem{Liddle:2003as}
A.R.~Liddle and S.M.~Leach, \emph{{How long before the end of inflation were
  observable perturbations produced?}},
  \href{https://doi.org/10.1103/PhysRevD.68.103503}{\emph{Phys. Rev. D}
  {\bfseries 68} (2003) 103503}
  [\href{https://arxiv.org/abs/astro-ph/0305263}{{\ttfamily
  astro-ph/0305263}}].

\bibitem{Planck:2018jri}
{\scshape Planck} collaboration, \emph{{Planck 2018 results. X. Constraints on
  inflation}}, \href{https://doi.org/10.1051/0004-6361/201833887}{\emph{Astron.
  Astrophys.} {\bfseries 641} (2020) A10}
  [\href{https://arxiv.org/abs/1807.06211}{{\ttfamily 1807.06211}}].

\bibitem{Ballesteros:2017fsr}
G.~Ballesteros and M.~Taoso, \emph{{Primordial black hole dark matter from
  single field inflation}},
  \href{https://doi.org/10.1103/PhysRevD.97.023501}{\emph{Phys. Rev. D}
  {\bfseries 97} (2018) 023501}
  [\href{https://arxiv.org/abs/1709.05565}{{\ttfamily 1709.05565}}].

\bibitem{Clesse:2015wea}
S.~Clesse and J.~Garc\'\i{}a-Bellido, \emph{{Massive Primordial Black Holes
  from Hybrid Inflation as Dark Matter and the seeds of Galaxies}},
  \href{https://doi.org/10.1103/PhysRevD.92.023524}{\emph{Phys. Rev. D}
  {\bfseries 92} (2015) 023524}
  [\href{https://arxiv.org/abs/1501.07565}{{\ttfamily 1501.07565}}].

\bibitem{Kinney:2005vj}
W.H.~Kinney, \emph{{Horizon crossing and inflation with large eta}},
  \href{https://doi.org/10.1103/PhysRevD.72.023515}{\emph{Phys. Rev. D}
  {\bfseries 72} (2005) 023515}
  [\href{https://arxiv.org/abs/gr-qc/0503017}{{\ttfamily gr-qc/0503017}}].

\bibitem{Motohashi:2017kbs}
H.~Motohashi and W.~Hu, \emph{{Primordial Black Holes and Slow-Roll
  Violation}}, \href{https://doi.org/10.1103/PhysRevD.96.063503}{\emph{Phys.
  Rev. D} {\bfseries 96} (2017) 063503}
  [\href{https://arxiv.org/abs/1706.06784}{{\ttfamily 1706.06784}}].

\bibitem{Dalianis:2018frf}
I.~Dalianis, A.~Kehagias and G.~Tringas, \emph{{Primordial black holes from
  \ensuremath{\alpha}-attractors}},
  \href{https://doi.org/10.1088/1475-7516/2019/01/037}{\emph{JCAP} {\bfseries
  01} (2019) 037} [\href{https://arxiv.org/abs/1805.09483}{{\ttfamily
  1805.09483}}].

\bibitem{Wu:2021mwy}
Y.-P.~Wu, E.~Pinetti, K.~Petraki and J.~Silk, \emph{{Baryogenesis from
  ultra-slow-roll inflation}},
  \href{https://doi.org/10.1007/JHEP01(2022)015}{\emph{JHEP} {\bfseries 01}
  (2022) 015} [\href{https://arxiv.org/abs/2109.00118}{{\ttfamily
  2109.00118}}].

\bibitem{Hawking:1975vcx}
S.W.~Hawking, \emph{{Particle Creation by Black Holes}},
  \href{https://doi.org/10.1007/BF02345020}{\emph{Commun. Math. Phys.}
  {\bfseries 43} (1975) 199}.

\bibitem{Auffinger:2022khh}
J.~Auffinger, \emph{{Primordial black hole constraints with Hawking radiation
  -- a review}},  \href{https://arxiv.org/abs/2206.02672}{{\ttfamily
  2206.02672}}.

\bibitem{Page:1976wx}
D.N.~Page and S.W.~Hawking, \emph{{Gamma rays from primordial black holes}},
  \href{https://doi.org/10.1086/154350}{\emph{Astrophys. J.} {\bfseries 206}
  (1976) 1}.

\bibitem{Carr:1976zz}
B.J.~Carr, \emph{{Some cosmological consequences of primordial black-hole
  evaporations}}, \href{https://doi.org/10.1086/154351}{\emph{Astrophys. J.}
  {\bfseries 206} (1976) 8}.

\bibitem{MacGibbon:1990zk}
J.H.~MacGibbon and B.R.~Webber, \emph{{Quark and gluon jet emission from
  primordial black holes: The instantaneous spectra}},
  \href{https://doi.org/10.1103/PhysRevD.41.3052}{\emph{Phys. Rev. D}
  {\bfseries 41} (1990) 3052}.

\bibitem{MacGibbon:1991tj}
J.H.~MacGibbon, \emph{{Quark and gluon jet emission from primordial black
  holes. 2. The Lifetime emission}},
  \href{https://doi.org/10.1103/PhysRevD.44.376}{\emph{Phys. Rev. D} {\bfseries
  44} (1991) 376}.

\bibitem{Teukolsky:1973ha}
S.A.~Teukolsky, \emph{{Perturbations of a rotating black hole. 1. Fundamental
  equations for gravitational electromagnetic and neutrino field
  perturbations}}, \href{https://doi.org/10.1086/152444}{\emph{Astrophys. J.}
  {\bfseries 185} (1973) 635}.

\bibitem{Bardeen:1973xb}
J.M.~Bardeen and W.H.~Press, \emph{{Radiation fields in the schwarzschild
  background}}, \href{https://doi.org/10.1063/1.1666175}{\emph{J. Math. Phys.}
  {\bfseries 14} (1973) 7}.

\bibitem{Page:1977um}
D.N.~Page, \emph{{Particle Emission Rates from a Black Hole. 3. Charged Leptons
  from a Nonrotating Hole}},
  \href{https://doi.org/10.1103/PhysRevD.16.2402}{\emph{Phys. Rev. D}
  {\bfseries 16} (1977) 2402}.

\bibitem{Kawasaki:1999na}
M.~Kawasaki, K.~Kohri and N.~Sugiyama, \emph{{Cosmological constraints on late
  time entropy production}},
  \href{https://doi.org/10.1103/PhysRevLett.82.4168}{\emph{Phys. Rev. Lett.}
  {\bfseries 82} (1999) 4168}
  [\href{https://arxiv.org/abs/astro-ph/9811437}{{\ttfamily
  astro-ph/9811437}}].

\bibitem{Kawasaki:2000en}
M.~Kawasaki, K.~Kohri and N.~Sugiyama, \emph{{MeV scale reheating temperature
  and thermalization of neutrino background}},
  \href{https://doi.org/10.1103/PhysRevD.62.023506}{\emph{Phys. Rev. D}
  {\bfseries 62} (2000) 023506}
  [\href{https://arxiv.org/abs/astro-ph/0002127}{{\ttfamily
  astro-ph/0002127}}].

\bibitem{Hannestad:2004px}
S.~Hannestad, \emph{{What is the lowest possible reheating temperature?}},
  \href{https://doi.org/10.1103/PhysRevD.70.043506}{\emph{Phys. Rev. D}
  {\bfseries 70} (2004) 043506}
  [\href{https://arxiv.org/abs/astro-ph/0403291}{{\ttfamily
  astro-ph/0403291}}].

\bibitem{Ichikawa:2005vw}
K.~Ichikawa, M.~Kawasaki and F.~Takahashi, \emph{{The Oscillation effects on
  thermalization of the neutrinos in the Universe with low reheating
  temperature}}, \href{https://doi.org/10.1103/PhysRevD.72.043522}{\emph{Phys.
  Rev. D} {\bfseries 72} (2005) 043522}
  [\href{https://arxiv.org/abs/astro-ph/0505395}{{\ttfamily
  astro-ph/0505395}}].

\bibitem{Domenech:2020ssp}
G.~Dom\`enech, C.~Lin and M.~Sasaki, \emph{{Gravitational wave constraints on
  the primordial black hole dominated early universe}},
  \href{https://doi.org/10.1088/1475-7516/2021/11/E01}{\emph{JCAP} {\bfseries
  04} (2021) 062} [\href{https://arxiv.org/abs/2012.08151}{{\ttfamily
  2012.08151}}].

\bibitem{Workman:2022ynf}
{\scshape Particle Data Group} collaboration, \emph{{Review of Particle
  Physics}}, \href{https://doi.org/10.1093/ptep/ptac097}{\emph{PTEP} {\bfseries
  2022} (2022) 083C01}.

\bibitem{Nakamura:1997sm}
T.~Nakamura, M.~Sasaki, T.~Tanaka and K.S.~Thorne, \emph{{Gravitational waves
  from coalescing black hole MACHO binaries}},
  \href{https://doi.org/10.1086/310886}{\emph{Astrophys. J. Lett.} {\bfseries
  487} (1997) L139} [\href{https://arxiv.org/abs/astro-ph/9708060}{{\ttfamily
  astro-ph/9708060}}].

\bibitem{Zagorac:2019ekv}
J.L.~Zagorac, R.~Easther and N.~Padmanabhan, \emph{{GUT-Scale Primordial Black
  Holes: Mergers and Gravitational Waves}},
  \href{https://doi.org/10.1088/1475-7516/2019/06/052}{\emph{JCAP} {\bfseries
  06} (2019) 052} [\href{https://arxiv.org/abs/1903.05053}{{\ttfamily
  1903.05053}}].

\bibitem{Bondi:1952ni}
H.~Bondi, \emph{{On spherically symmetrical accretion}},
  \href{https://doi.org/10.1093/mnras/112.2.195}{\emph{Mon. Not. Roy. Astron.
  Soc.} {\bfseries 112} (1952) 195}.

\bibitem{1966AZh....43..758Z}
Y.B.~{Zel'dovich} and I.D.~{Novikov}, \emph{{The Hypothesis of Cores Retarded
  during Expansion and the Hot Cosmological Model}}, {\emph{Astron. Rep.}
  {\bfseries 43} (1966) 758}.

\bibitem{Planck:2018vyg}
{\scshape Planck} collaboration, \emph{{Planck 2018 results. VI. Cosmological
  parameters}},
  \href{https://doi.org/10.1051/0004-6361/201833910}{\emph{Astron. Astrophys.}
  {\bfseries 641} (2020) A6}
  [\href{https://arxiv.org/abs/1807.06209}{{\ttfamily 1807.06209}}].

\bibitem{Mangano:2001iu}
G.~Mangano, G.~Miele, S.~Pastor and M.~Peloso, \emph{{A Precision calculation
  of the effective number of cosmological neutrinos}},
  \href{https://doi.org/10.1016/S0370-2693(02)01622-2}{\emph{Phys. Lett. B}
  {\bfseries 534} (2002) 8}
  [\href{https://arxiv.org/abs/astro-ph/0111408}{{\ttfamily
  astro-ph/0111408}}].

\bibitem{deSalas:2016ztq}
P.F.~de~Salas and S.~Pastor, \emph{{Relic neutrino decoupling with flavour
  oscillations revisited}},
  \href{https://doi.org/10.1088/1475-7516/2016/07/051}{\emph{JCAP} {\bfseries
  07} (2016) 051} [\href{https://arxiv.org/abs/1606.06986}{{\ttfamily
  1606.06986}}].

\bibitem{Akita:2020szl}
K.~Akita and M.~Yamaguchi, \emph{{A precision calculation of relic neutrino
  decoupling}},
  \href{https://doi.org/10.1088/1475-7516/2020/08/012}{\emph{JCAP} {\bfseries
  08} (2020) 012} [\href{https://arxiv.org/abs/2005.07047}{{\ttfamily
  2005.07047}}].

\bibitem{Bennett:2020zkv}
J.J.~Bennett, G.~Buldgen, P.F.~De~Salas, M.~Drewes, S.~Gariazzo, S.~Pastor
  et~al., \emph{{Towards a precision calculation of $N_{\rm eff}$ in the
  Standard Model II: Neutrino decoupling in the presence of flavour
  oscillations and finite-temperature QED}},
  \href{https://doi.org/10.1088/1475-7516/2021/04/073}{\emph{JCAP} {\bfseries
  04} (2021) 073} [\href{https://arxiv.org/abs/2012.02726}{{\ttfamily
  2012.02726}}].

\bibitem{Papanikolaou:2020qtd}
T.~Papanikolaou, V.~Vennin and D.~Langlois, \emph{{Gravitational waves from a
  universe filled with primordial black holes}},
  \href{https://doi.org/10.1088/1475-7516/2021/03/053}{\emph{JCAP} {\bfseries
  03} (2021) 053} [\href{https://arxiv.org/abs/2010.11573}{{\ttfamily
  2010.11573}}].

\bibitem{Inomata:2020lmk}
K.~Inomata, M.~Kawasaki, K.~Mukaida, T.~Terada and T.T.~Yanagida,
  \emph{{Gravitational Wave Production right after a Primordial Black Hole
  Evaporation}}, \href{https://doi.org/10.1103/PhysRevD.101.123533}{\emph{Phys.
  Rev. D} {\bfseries 101} (2020) 123533}
  [\href{https://arxiv.org/abs/2003.10455}{{\ttfamily 2003.10455}}].

\bibitem{Dolgov:2011cq}
A.D.~Dolgov and D.~Ejlli, \emph{{Relic gravitational waves from light
  primordial black holes}},
  \href{https://doi.org/10.1103/PhysRevD.84.024028}{\emph{Phys. Rev. D}
  {\bfseries 84} (2011) 024028}
  [\href{https://arxiv.org/abs/1105.2303}{{\ttfamily 1105.2303}}].

\bibitem{Caprini:2018mtu}
C.~Caprini and D.G.~Figueroa, \emph{{Cosmological Backgrounds of Gravitational
  Waves}}, \href{https://doi.org/10.1088/1361-6382/aac608}{\emph{Class. Quant.
  Grav.} {\bfseries 35} (2018) 163001}
  [\href{https://arxiv.org/abs/1801.04268}{{\ttfamily 1801.04268}}].

\bibitem{Ando:2018qdb}
K.~Ando, K.~Inomata and M.~Kawasaki, \emph{{Primordial black holes and
  uncertainties in the choice of the window function}},
  \href{https://doi.org/10.1103/PhysRevD.97.103528}{\emph{Phys. Rev. D}
  {\bfseries 97} (2018) 103528}
  [\href{https://arxiv.org/abs/1802.06393}{{\ttfamily 1802.06393}}].

\bibitem{Inomata:2019ivs}
K.~Inomata, K.~Kohri, T.~Nakama and T.~Terada, \emph{{Enhancement of
  Gravitational Waves Induced by Scalar Perturbations due to a Sudden
  Transition from an Early Matter Era to the Radiation Era}},
  \href{https://doi.org/10.1103/PhysRevD.100.043532}{\emph{Phys. Rev. D}
  {\bfseries 100} (2019) 043532}
  [\href{https://arxiv.org/abs/1904.12879}{{\ttfamily 1904.12879}}].

\bibitem{Passaglia:2021jla}
S.~Passaglia and M.~Sasaki, \emph{{Primordial black holes from CDM isocurvature
  perturbations}},
  \href{https://doi.org/10.1103/PhysRevD.105.103530}{\emph{Phys. Rev. D}
  {\bfseries 105} (2022) 103530}
  [\href{https://arxiv.org/abs/2109.12824}{{\ttfamily 2109.12824}}].

\bibitem{Domenech:2021and}
G.~Dom\`enech, S.~Passaglia and S.~Renaux-Petel, \emph{{Gravitational waves
  from dark matter isocurvature}},
  \href{https://doi.org/10.1088/1475-7516/2022/03/023}{\emph{JCAP} {\bfseries
  03} (2022) 023} [\href{https://arxiv.org/abs/2112.10163}{{\ttfamily
  2112.10163}}].

\bibitem{Kim:1979if}
J.E.~Kim, \emph{{Weak Interaction Singlet and Strong CP Invariance}},
  \href{https://doi.org/10.1103/PhysRevLett.43.103}{\emph{Phys. Rev. Lett.}
  {\bfseries 43} (1979) 103}.

\bibitem{Shifman:1979if}
M.A.~Shifman, A.I.~Vainshtein and V.I.~Zakharov, \emph{{Can Confinement Ensure
  Natural CP Invariance of Strong Interactions?}},
  \href{https://doi.org/10.1016/0550-3213(80)90209-6}{\emph{Nucl. Phys. B}
  {\bfseries 166} (1980) 493}.

\bibitem{Zhitnitsky:1980tq}
A.R.~Zhitnitsky, \emph{{On Possible Suppression of the Axion Hadron
  Interactions. (In Russian)}}, {\emph{Sov. J. Nucl. Phys.} {\bfseries 31}
  (1980) 260}.

\bibitem{Dine:1981rt}
M.~Dine, W.~Fischler and M.~Srednicki, \emph{{A Simple Solution to the Strong
  CP Problem with a Harmless Axion}},
  \href{https://doi.org/10.1016/0370-2693(81)90590-6}{\emph{Phys. Lett. B}
  {\bfseries 104} (1981) 199}.

\bibitem{DiLuzio:2020wdo}
L.~Di~Luzio, M.~Giannotti, E.~Nardi and L.~Visinelli, \emph{{The landscape of
  QCD axion models}},
  \href{https://doi.org/10.1016/j.physrep.2020.06.002}{\emph{Phys. Rept.}
  {\bfseries 870} (2020) 1} [\href{https://arxiv.org/abs/2003.01100}{{\ttfamily
  2003.01100}}].

\bibitem{Borsanyi:2016ksw}
S.~Borsanyi et~al., \emph{{Calculation of the axion mass based on
  high-temperature lattice quantum chromodynamics}},
  \href{https://doi.org/10.1038/nature20115}{\emph{Nature} {\bfseries 539}
  (2016) 69} [\href{https://arxiv.org/abs/1606.07494}{{\ttfamily 1606.07494}}].

\bibitem{Petreczky:2016vrs}
P.~Petreczky, H.-P.~Schadler and S.~Sharma, \emph{{The topological
  susceptibility in finite temperature QCD and axion cosmology}},
  \href{https://doi.org/10.1016/j.physletb.2016.09.063}{\emph{Phys. Lett. B}
  {\bfseries 762} (2016) 498}
  [\href{https://arxiv.org/abs/1606.03145}{{\ttfamily 1606.03145}}].

\bibitem{Gross:1980br}
D.J.~Gross, R.D.~Pisarski and L.G.~Yaffe, \emph{{QCD and Instantons at Finite
  Temperature}}, \href{https://doi.org/10.1103/RevModPhys.53.43}{\emph{Rev.
  Mod. Phys.} {\bfseries 53} (1981) 43}.

\bibitem{Salvio:2013iaa}
A.~Salvio, A.~Strumia and W.~Xue, \emph{{Thermal axion production}},
  \href{https://doi.org/10.1088/1475-7516/2014/01/011}{\emph{JCAP} {\bfseries
  01} (2014) 011} [\href{https://arxiv.org/abs/1310.6982}{{\ttfamily
  1310.6982}}].

\bibitem{Baumann:2016wac}
D.~Baumann, D.~Green and B.~Wallisch, \emph{{New Target for Cosmic Axion
  Searches}}, \href{https://doi.org/10.1103/PhysRevLett.117.171301}{\emph{Phys.
  Rev. Lett.} {\bfseries 117} (2016) 171301}
  [\href{https://arxiv.org/abs/1604.08614}{{\ttfamily 1604.08614}}].

\bibitem{Caloni:2022uya}
L.~Caloni, M.~Gerbino, M.~Lattanzi and L.~Visinelli, \emph{{Novel cosmological
  bounds on thermally-produced axion-like particles}},
  \href{https://doi.org/10.1088/1475-7516/2022/09/021}{\emph{JCAP} {\bfseries
  09} (2022) 021} [\href{https://arxiv.org/abs/2205.01637}{{\ttfamily
  2205.01637}}].

\bibitem{CMB-S4:2016ple}
{\scshape CMB-S4} collaboration, \emph{{CMB-S4 Science Book, First Edition}},
  \href{https://arxiv.org/abs/1610.02743}{{\ttfamily 1610.02743}}.

\bibitem{Abazajian:2016hbv}
K.N.~Abazajian and M.~Kaplinghat, \emph{{Neutrino Physics from the Cosmic
  Microwave Background and Large-Scale Structure}},
  \href{https://doi.org/10.1146/annurev-nucl-102014-021908}{\emph{Ann. Rev.
  Nucl. Part. Sci.} {\bfseries 66} (2016) 401}.

\bibitem{Abazajian:2019eic}
K.~Abazajian et~al., \emph{{CMB-S4 Science Case, Reference Design, and Project
  Plan}},  \href{https://arxiv.org/abs/1907.04473}{{\ttfamily 1907.04473}}.

\bibitem{Grin:2007yg}
D.~Grin, T.L.~Smith and M.~Kamionkowski, \emph{{Axion constraints in
  non-standard thermal histories}},
  \href{https://doi.org/10.1103/PhysRevD.77.085020}{\emph{Phys. Rev. D}
  {\bfseries 77} (2008) 085020}
  [\href{https://arxiv.org/abs/0711.1352}{{\ttfamily 0711.1352}}].

\bibitem{Chang:1993gm}
S.~Chang and K.~Choi, \emph{{Hadronic axion window and the big bang
  nucleosynthesis}},
  \href{https://doi.org/10.1016/0370-2693(93)90656-3}{\emph{Phys. Lett. B}
  {\bfseries 316} (1993) 51}
  [\href{https://arxiv.org/abs/hep-ph/9306216}{{\ttfamily hep-ph/9306216}}].

\bibitem{Hannestad:2005df}
S.~Hannestad, A.~Mirizzi and G.~Raffelt, \emph{{New cosmological mass limit on
  thermal relic axions}},
  \href{https://doi.org/10.1088/1475-7516/2005/07/002}{\emph{JCAP} {\bfseries
  07} (2005) 002} [\href{https://arxiv.org/abs/hep-ph/0504059}{{\ttfamily
  hep-ph/0504059}}].

\bibitem{Giare:2020vzo}
W.~Giar\`e, E.~Di~Valentino, A.~Melchiorri and O.~Mena, \emph{{New cosmological
  bounds on hot relics: axions and neutrinos}},
  \href{https://doi.org/10.1093/mnras/stab1442}{\emph{Mon. Not. Roy. Astron.
  Soc.} {\bfseries 505} (2021) 2703}
  [\href{https://arxiv.org/abs/2011.14704}{{\ttfamily 2011.14704}}].

\bibitem{Turner:1986tb}
M.S.~Turner, \emph{{Thermal Production of Not SO Invisible Axions in the Early
  Universe}}, \href{https://doi.org/10.1103/PhysRevLett.59.2489}{\emph{Phys.
  Rev. Lett.} {\bfseries 59} (1987) 2489}.

\bibitem{Graf:2010tv}
P.~Graf and F.D.~Steffen, \emph{{Thermal axion production in the primordial
  quark-gluon plasma}},
  \href{https://doi.org/10.1103/PhysRevD.83.075011}{\emph{Phys. Rev. D}
  {\bfseries 83} (2011) 075011}
  [\href{https://arxiv.org/abs/1008.4528}{{\ttfamily 1008.4528}}].

\bibitem{Linde:1985yf}
A.D.~Linde, \emph{{Generation of Isothermal Density Perturbations in the
  Inflationary Universe}},
  \href{https://doi.org/10.1016/0370-2693(85)90436-8}{\emph{Phys. Lett. B}
  {\bfseries 158} (1985) 375}.

\bibitem{Seckel:1985tj}
D.~Seckel and M.S.~Turner, \emph{{Isothermal Density Perturbations in an Axion
  Dominated Inflationary Universe}},
  \href{https://doi.org/10.1103/PhysRevD.32.3178}{\emph{Phys. Rev. D}
  {\bfseries 32} (1985) 3178}.

\bibitem{Turner:1990uz}
M.S.~Turner and F.~Wilczek, \emph{{Inflationary axion cosmology}},
  \href{https://doi.org/10.1103/PhysRevLett.66.5}{\emph{Phys. Rev. Lett.}
  {\bfseries 66} (1991) 5}.

\bibitem{Linde:1991km}
A.D.~Linde, \emph{{Axions in inflationary cosmology}},
  \href{https://doi.org/10.1016/0370-2693(91)90130-I}{\emph{Phys. Lett. B}
  {\bfseries 259} (1991) 38}.

\bibitem{Beltran:2006sq}
M.~Beltran, J.~Garcia-Bellido and J.~Lesgourgues, \emph{{Isocurvature bounds on
  axions revisited}},
  \href{https://doi.org/10.1103/PhysRevD.75.103507}{\emph{Phys. Rev. D}
  {\bfseries 75} (2007) 103507}
  [\href{https://arxiv.org/abs/hep-ph/0606107}{{\ttfamily hep-ph/0606107}}].

\bibitem{Turner:1985si}
M.S.~Turner, \emph{{Cosmic and Local Mass Density of Invisible Axions}},
  \href{https://doi.org/10.1103/PhysRevD.33.889}{\emph{Phys. Rev. D} {\bfseries
  33} (1986) 889}.

\bibitem{Strobl:1994wk}
K.~Strobl and T.J.~Weiler, \emph{{Anharmonic evolution of the cosmic axion
  density spectrum}},
  \href{https://doi.org/10.1103/PhysRevD.50.7690}{\emph{Phys. Rev. D}
  {\bfseries 50} (1994) 7690}
  [\href{https://arxiv.org/abs/astro-ph/9405028}{{\ttfamily
  astro-ph/9405028}}].

\bibitem{Bae:2008ue}
K.J.~Bae, J.-H.~Huh and J.E.~Kim, \emph{{Update of axion CDM energy}},
  \href{https://doi.org/10.1088/1475-7516/2008/09/005}{\emph{JCAP} {\bfseries
  09} (2008) 005} [\href{https://arxiv.org/abs/0806.0497}{{\ttfamily
  0806.0497}}].

\bibitem{Visinelli:2009zm}
L.~Visinelli and P.~Gondolo, \emph{{Dark Matter Axions Revisited}},
  \href{https://doi.org/10.1103/PhysRevD.80.035024}{\emph{Phys. Rev. D}
  {\bfseries 80} (2009) 035024}
  [\href{https://arxiv.org/abs/0903.4377}{{\ttfamily 0903.4377}}].

\bibitem{Graham:2018jyp}
P.W.~Graham and A.~Scherlis, \emph{{Stochastic axion scenario}},
  \href{https://doi.org/10.1103/PhysRevD.98.035017}{\emph{Phys. Rev. D}
  {\bfseries 98} (2018) 035017}
  [\href{https://arxiv.org/abs/1805.07362}{{\ttfamily 1805.07362}}].

\bibitem{Takahashi:2018tdu}
F.~Takahashi, W.~Yin and A.H.~Guth, \emph{{QCD axion window and low-scale
  inflation}}, \href{https://doi.org/10.1103/PhysRevD.98.015042}{\emph{Phys.
  Rev. D} {\bfseries 98} (2018) 015042}
  [\href{https://arxiv.org/abs/1805.08763}{{\ttfamily 1805.08763}}].

\bibitem{Schmitz:2018nhb}
K.~Schmitz and T.T.~Yanagida, \emph{{Axion Isocurvature Perturbations in
  Low-Scale Models of Hybrid Inflation}},
  \href{https://doi.org/10.1103/PhysRevD.98.075003}{\emph{Phys. Rev. D}
  {\bfseries 98} (2018) 075003}
  [\href{https://arxiv.org/abs/1806.06056}{{\ttfamily 1806.06056}}].

\bibitem{Tenkanen:2019xzn}
T.~Tenkanen and L.~Visinelli, \emph{{Axion dark matter from Higgs inflation
  with an intermediate $H_*$}},
  \href{https://doi.org/10.1088/1475-7516/2019/08/033}{\emph{JCAP} {\bfseries
  08} (2019) 033} [\href{https://arxiv.org/abs/1906.11837}{{\ttfamily
  1906.11837}}].

\bibitem{Odintsov:2019mlf}
S.D.~Odintsov and V.K.~Oikonomou, \emph{{$f(R)$ Gravity Inflation with
  String-Corrected Axion Dark Matter}},
  \href{https://doi.org/10.1103/PhysRevD.99.064049}{\emph{Phys. Rev. D}
  {\bfseries 99} (2019) 064049}
  [\href{https://arxiv.org/abs/1901.05363}{{\ttfamily 1901.05363}}].

\bibitem{Odintsov:2019evb}
S.D.~Odintsov and V.K.~Oikonomou, \emph{{Unification of Inflation with Dark
  Energy in $f(R)$ Gravity and Axion Dark Matter}},
  \href{https://doi.org/10.1103/PhysRevD.99.104070}{\emph{Phys. Rev. D}
  {\bfseries 99} (2019) 104070}
  [\href{https://arxiv.org/abs/1905.03496}{{\ttfamily 1905.03496}}].

\bibitem{Lambiase:2022ucu}
G.~Lambiase, L.~Mastrototaro and L.~Visinelli, \emph{{Gravitational waves and
  neutrino oscillations in Chern-Simons axion gravity}},
  \href{https://doi.org/10.1088/1475-7516/2023/01/011}{\emph{JCAP} {\bfseries
  01} (2023) 011} [\href{https://arxiv.org/abs/2207.08067}{{\ttfamily
  2207.08067}}].

\bibitem{Hoof:2018ieb}
S.~Hoof, F.~Kahlhoefer, P.~Scott, C.~Weniger and M.~White, \emph{{Axion global
  fits with Peccei-Quinn symmetry breaking before inflation using GAMBIT}},
  \href{https://doi.org/10.1007/JHEP03(2019)191}{\emph{JHEP} {\bfseries 03}
  (2019) 191} [\href{https://arxiv.org/abs/1810.07192}{{\ttfamily
  1810.07192}}].

\bibitem{Seto:2001qf}
N.~Seto, S.~Kawamura and T.~Nakamura, \emph{{Possibility of direct measurement
  of the acceleration of the universe using 0.1-Hz band laser interferometer
  gravitational wave antenna in space}},
  \href{https://doi.org/10.1103/PhysRevLett.87.221103}{\emph{Phys. Rev. Lett.}
  {\bfseries 87} (2001) 221103}
  [\href{https://arxiv.org/abs/astro-ph/0108011}{{\ttfamily
  astro-ph/0108011}}].

\bibitem{Yagi:2011wg}
K.~Yagi and N.~Seto, \emph{{Detector configuration of DECIGO/BBO and
  identification of cosmological neutron-star binaries}},
  \href{https://doi.org/10.1103/PhysRevD.83.044011}{\emph{Phys. Rev. D}
  {\bfseries 83} (2011) 044011}
  [\href{https://arxiv.org/abs/1101.3940}{{\ttfamily 1101.3940}}].

\bibitem{Maggiore:2019uih}
M.~Maggiore et~al., \emph{{Science Case for the Einstein Telescope}},
  \href{https://doi.org/10.1088/1475-7516/2020/03/050}{\emph{JCAP} {\bfseries
  03} (2020) 050} [\href{https://arxiv.org/abs/1912.02622}{{\ttfamily
  1912.02622}}].

\bibitem{2017arXiv170200786A}
P.~{Amaro-Seoane} et~al., \emph{{Laser Interferometer Space Antenna}},
  \href{https://arxiv.org/abs/1702.00786}{{\ttfamily 1702.00786}}.

\bibitem{1977SvAL....3..110Z}
I.B.~{Zeldovich}, A.A.~{Starobinskii}, M.I.~{Khlopov} and V.M.~{Chechetkin},
  \emph{{Primordial black holes and the deuterium problem}}, {\emph{Soviet
  Astronomy Letters} {\bfseries 3} (1977) 110}.

\bibitem{Carr:1994ar}
B.J.~Carr, J.H.~Gilbert and J.E.~Lidsey, \emph{{Black hole relics and
  inflation: Limits on blue perturbation spectra}},
  \href{https://doi.org/10.1103/PhysRevD.50.4853}{\emph{Phys. Rev. D}
  {\bfseries 50} (1994) 4853}
  [\href{https://arxiv.org/abs/astro-ph/9405027}{{\ttfamily
  astro-ph/9405027}}].

\bibitem{Hogan:1988mp}
C.J.~Hogan and M.J.~Rees, \emph{{Axion Miniclusters}},
  \href{https://doi.org/10.1016/0370-2693(88)91655-3}{\emph{Phys. Lett. B}
  {\bfseries 205} (1988) 228}.

\bibitem{Blinov:2019jqc}
N.~Blinov, M.J.~Dolan and P.~Draper, \emph{{Imprints of the Early Universe on
  Axion Dark Matter Substructure}},
  \href{https://doi.org/10.1103/PhysRevD.101.035002}{\emph{Phys. Rev. D}
  {\bfseries 101} (2020) 035002}
  [\href{https://arxiv.org/abs/1911.07853}{{\ttfamily 1911.07853}}].

\bibitem{Erickcek:2011us}
A.L.~Erickcek and K.~Sigurdson, \emph{{Reheating Effects in the Matter Power
  Spectrum and Implications for Substructure}},
  \href{https://doi.org/10.1103/PhysRevD.84.083503}{\emph{Phys. Rev. D}
  {\bfseries 84} (2011) 083503}
  [\href{https://arxiv.org/abs/1106.0536}{{\ttfamily 1106.0536}}].

\bibitem{Klaer:2017ond}
V.B..~Klaer and G.D.~Moore, \emph{{The dark-matter axion mass}},
  \href{https://doi.org/10.1088/1475-7516/2017/11/049}{\emph{JCAP} {\bfseries
  11} (2017) 049} [\href{https://arxiv.org/abs/1708.07521}{{\ttfamily
  1708.07521}}].

\bibitem{Gorghetto:2018myk}
M.~Gorghetto, E.~Hardy and G.~Villadoro, \emph{{Axions from Strings: the
  Attractive Solution}},
  \href{https://doi.org/10.1007/JHEP07(2018)151}{\emph{JHEP} {\bfseries 07}
  (2018) 151} [\href{https://arxiv.org/abs/1806.04677}{{\ttfamily
  1806.04677}}].

\bibitem{Vaquero:2018tib}
A.~Vaquero, J.~Redondo and J.~Stadler, \emph{{Early seeds of axion
  miniclusters}},
  \href{https://doi.org/10.1088/1475-7516/2019/04/012}{\emph{JCAP} {\bfseries
  04} (2019) 012} [\href{https://arxiv.org/abs/1809.09241}{{\ttfamily
  1809.09241}}].

\bibitem{Gorghetto:2020qws}
M.~Gorghetto, E.~Hardy and G.~Villadoro, \emph{{More axions from strings}},
  \href{https://doi.org/10.21468/SciPostPhys.10.2.050}{\emph{SciPost Phys.}
  {\bfseries 10} (2021) 050}
  [\href{https://arxiv.org/abs/2007.04990}{{\ttfamily 2007.04990}}].

\bibitem{Buschmann:2021sdq}
M.~Buschmann, J.W.~Foster, A.~Hook, A.~Peterson, D.E.~Willcox, W.~Zhang et~al.,
  \emph{{Dark matter from axion strings with adaptive mesh refinement}},
  \href{https://doi.org/10.1038/s41467-022-28669-y}{\emph{Nature Commun.}
  {\bfseries 13} (2022) 1049}
  [\href{https://arxiv.org/abs/2108.05368}{{\ttfamily 2108.05368}}].

\bibitem{Eggemeier:2022hqa}
B.~Eggemeier, C.A.J.~O'Hare, G.~Pierobon, J.~Redondo and Y.Y.Y.~Wong,
  \emph{{Axion minivoids and implications for direct detection}},
  \href{https://arxiv.org/abs/2212.00560}{{\ttfamily 2212.00560}}.

\bibitem{Yamaguchi:1999dy}
M.~Yamaguchi, J.~Yokoyama and M.~Kawasaki, \emph{{Evolution of a global string
  network in a matter dominated universe}},
  \href{https://doi.org/10.1103/PhysRevD.61.061301}{\emph{Phys. Rev. D}
  {\bfseries 61} (2000) 061301}
  [\href{https://arxiv.org/abs/hep-ph/9910352}{{\ttfamily hep-ph/9910352}}].

\bibitem{Lee:2020wfn}
V.S.H.~Lee, A.~Mitridate, T.~Trickle and K.M.~Zurek, \emph{{Probing Small-Scale
  Power Spectra with Pulsar Timing Arrays}},
  \href{https://doi.org/10.1007/JHEP06(2021)028}{\emph{JHEP} {\bfseries 06}
  (2021) 028} [\href{https://arxiv.org/abs/2012.09857}{{\ttfamily
  2012.09857}}].

\bibitem{Dror:2019twh}
J.A.~Dror, H.~Ramani, T.~Trickle and K.M.~Zurek, \emph{{Pulsar Timing Probes of
  Primordial Black Holes and Subhalos}},
  \href{https://doi.org/10.1103/PhysRevD.100.023003}{\emph{Phys. Rev. D}
  {\bfseries 100} (2019) 023003}
  [\href{https://arxiv.org/abs/1901.04490}{{\ttfamily 1901.04490}}].

\bibitem{Dokuchaev:2017psd}
V.I.~Dokuchaev, Y.N.~Eroshenko and I.I.~Tkachev, \emph{{Destruction of axion
  miniclusters in the Galaxy}},
  \href{https://doi.org/10.1134/S1063776117080039}{\emph{J. Exp. Theor. Phys.}
  {\bfseries 125} (2017) 434}
  [\href{https://arxiv.org/abs/1710.09586}{{\ttfamily 1710.09586}}].

\bibitem{Kavanagh:2020gcy}
B.J.~Kavanagh, T.D.P.~Edwards, L.~Visinelli and C.~Weniger, \emph{{Stellar
  disruption of axion miniclusters in the Milky~Way}},
  \href{https://doi.org/10.1103/PhysRevD.104.063038}{\emph{Phys. Rev. D}
  {\bfseries 104} (2021) 063038}
  [\href{https://arxiv.org/abs/2011.05377}{{\ttfamily 2011.05377}}].

\bibitem{Edwards:2020afl}
T.D.P.~Edwards, B.J.~Kavanagh, L.~Visinelli and C.~Weniger, \emph{{Transient
  Radio Signatures from Neutron Star Encounters with QCD Axion Miniclusters}},
  \href{https://doi.org/10.1103/PhysRevLett.127.131103}{\emph{Phys. Rev. Lett.}
  {\bfseries 127} (2021) 131103}
  [\href{https://arxiv.org/abs/2011.05378}{{\ttfamily 2011.05378}}].

\bibitem{Shen:2022ltx}
X.~Shen, H.~Xiao, P.F.~Hopkins and K.M.~Zurek, \emph{{Disruption of Dark Matter
  Minihaloes in the Milky Way environment: Implications for Axion Miniclusters
  and Early Matter Domination}},
  \href{https://arxiv.org/abs/2207.11276}{{\ttfamily 2207.11276}}.

\bibitem{Dai:2019lud}
L.~Dai and J.~Miralda-Escud\'e, \emph{{Gravitational Lensing Signatures of
  Axion Dark Matter Minihalos in Highly Magnified Stars}},
  \href{https://doi.org/10.3847/1538-3881/ab5e83}{\emph{Astron. J.} {\bfseries
  159} (2020) 49} [\href{https://arxiv.org/abs/1908.01773}{{\ttfamily
  1908.01773}}].

\bibitem{Berlin:2019ahk}
A.~Berlin, R.T.~D'Agnolo, S.A.R.~Ellis, C.~Nantista, J.~Neilson, P.~Schuster
  et~al., \emph{{Axion Dark Matter Detection by Superconducting Resonant
  Frequency Conversion}},
  \href{https://doi.org/10.1007/JHEP07(2020)088}{\emph{JHEP} {\bfseries 07}
  (2020) 088} [\href{https://arxiv.org/abs/1912.11048}{{\ttfamily
  1912.11048}}].

\bibitem{Berlin:2020vrk}
A.~Berlin, R.T.~D'Agnolo, S.A.R.~Ellis and K.~Zhou, \emph{{Heterodyne broadband
  detection of axion dark matter}},
  \href{https://doi.org/10.1103/PhysRevD.104.L111701}{\emph{Phys. Rev. D}
  {\bfseries 104} (2021) L111701}
  [\href{https://arxiv.org/abs/2007.15656}{{\ttfamily 2007.15656}}].

\bibitem{Berlin:2022hfx}
A.~Berlin et~al., \emph{{Searches for New Particles, Dark Matter, and
  Gravitational Waves with SRF Cavities}},
  \href{https://arxiv.org/abs/2203.12714}{{\ttfamily 2203.12714}}.

\bibitem{Niemeyer:1997mt}
J.C.~Niemeyer and K.~Jedamzik, \emph{{Near-critical gravitational collapse and
  the initial mass function of primordial black holes}},
  \href{https://doi.org/10.1103/PhysRevLett.80.5481}{\emph{Phys. Rev. Lett.}
  {\bfseries 80} (1998) 5481}
  [\href{https://arxiv.org/abs/astro-ph/9709072}{{\ttfamily
  astro-ph/9709072}}].

\bibitem{Koike:1995jm}
T.~Koike, T.~Hara and S.~Adachi, \emph{{Critical behavior in gravitational
  collapse of radiation fluid: A Renormalization group (linear perturbation)
  analysis}}, \href{https://doi.org/10.1103/PhysRevLett.74.5170}{\emph{Phys.
  Rev. Lett.} {\bfseries 74} (1995) 5170}
  [\href{https://arxiv.org/abs/gr-qc/9503007}{{\ttfamily gr-qc/9503007}}].

\bibitem{Green:2004wb}
A.M.~Green, A.R.~Liddle, K.A.~Malik and M.~Sasaki, \emph{{A New calculation of
  the mass fraction of primordial black holes}},
  \href{https://doi.org/10.1103/PhysRevD.70.041502}{\emph{Phys. Rev. D}
  {\bfseries 70} (2004) 041502}
  [\href{https://arxiv.org/abs/astro-ph/0403181}{{\ttfamily
  astro-ph/0403181}}].

\bibitem{Green:2016xgy}
A.M.~Green, \emph{{Microlensing and dynamical constraints on primordial black
  hole dark matter with an extended mass function}},
  \href{https://doi.org/10.1103/PhysRevD.94.063530}{\emph{Phys. Rev. D}
  {\bfseries 94} (2016) 063530}
  [\href{https://arxiv.org/abs/1609.01143}{{\ttfamily 1609.01143}}].

\bibitem{Kristiano:2022maq}
J.~Kristiano and J.~Yokoyama, \emph{{Ruling Out Primordial Black Hole Formation
  From Single-Field Inflation}},
  \href{https://arxiv.org/abs/2211.03395}{{\ttfamily 2211.03395}}.

\bibitem{Inomata:2022yte}
K.~Inomata, M.~Braglia and X.~Chen, \emph{{Questions on calculation of
  primordial power spectrum with large spikes: the resonance model case}},
  \href{https://arxiv.org/abs/2211.02586}{{\ttfamily 2211.02586}}.

\bibitem{Young:2013oia}
S.~Young and C.T.~Byrnes, \emph{{Primordial black holes in non-Gaussian
  regimes}}, \href{https://doi.org/10.1088/1475-7516/2013/08/052}{\emph{JCAP}
  {\bfseries 08} (2013) 052} [\href{https://arxiv.org/abs/1307.4995}{{\ttfamily
  1307.4995}}].

\bibitem{Young:2015kda}
S.~Young and C.T.~Byrnes, \emph{{Signatures of non-gaussianity in the
  isocurvature modes of primordial black hole dark matter}},
  \href{https://doi.org/10.1088/1475-7516/2015/04/034}{\emph{JCAP} {\bfseries
  04} (2015) 034} [\href{https://arxiv.org/abs/1503.01505}{{\ttfamily
  1503.01505}}].

\bibitem{Gelmini:2006pq}
G.~Gelmini, P.~Gondolo, A.~Soldatenko and C.E.~Yaguna, \emph{{The Effect of a
  late decaying scalar on the neutralino relic density}},
  \href{https://doi.org/10.1103/PhysRevD.74.083514}{\emph{Phys. Rev. D}
  {\bfseries 74} (2006) 083514}
  [\href{https://arxiv.org/abs/hep-ph/0605016}{{\ttfamily hep-ph/0605016}}].

\bibitem{Chung:1998zb}
D.J.H.~Chung, E.W.~Kolb and A.~Riotto, \emph{{Superheavy dark matter}},
  \href{https://doi.org/10.1103/PhysRevD.59.023501}{\emph{Phys. Rev. D}
  {\bfseries 59} (1998) 023501}
  [\href{https://arxiv.org/abs/hep-ph/9802238}{{\ttfamily hep-ph/9802238}}].

\bibitem{Lyth:1995ka}
D.H.~Lyth and E.D.~Stewart, \emph{{Thermal inflation and the moduli problem}},
  \href{https://doi.org/10.1103/PhysRevD.53.1784}{\emph{Phys. Rev. D}
  {\bfseries 53} (1996) 1784}
  [\href{https://arxiv.org/abs/hep-ph/9510204}{{\ttfamily hep-ph/9510204}}].

\bibitem{Coleman:2003hs}
T.S.~Coleman and M.~Roos, \emph{{Effective degrees of freedom during the
  radiation era}},
  \href{https://doi.org/10.1103/PhysRevD.68.027702}{\emph{Phys. Rev. D}
  {\bfseries 68} (2003) 027702}
  [\href{https://arxiv.org/abs/astro-ph/0304281}{{\ttfamily
  astro-ph/0304281}}].

\bibitem{Husdal:2016haj}
L.~Husdal, \emph{{On Effective Degrees of Freedom in the Early Universe}},
  \href{https://doi.org/10.3390/galaxies4040078}{\emph{Galaxies} {\bfseries 4}
  (2016) 78} [\href{https://arxiv.org/abs/1609.04979}{{\ttfamily 1609.04979}}].

\bibitem{Wantz:2009it}
O.~Wantz and E.P.S.~Shellard, \emph{{Axion Cosmology Revisited}},
  \href{https://doi.org/10.1103/PhysRevD.82.123508}{\emph{Phys. Rev. D}
  {\bfseries 82} (2010) 123508}
  [\href{https://arxiv.org/abs/0910.1066}{{\ttfamily 0910.1066}}].

\bibitem{Kolb:1993hw}
E.W.~Kolb and I.I.~Tkachev, \emph{{Nonlinear axion dynamics and formation of
  cosmological pseudosolitons}},
  \href{https://doi.org/10.1103/PhysRevD.49.5040}{\emph{Phys. Rev. D}
  {\bfseries 49} (1994) 5040}
  [\href{https://arxiv.org/abs/astro-ph/9311037}{{\ttfamily
  astro-ph/9311037}}].

\end{thebibliography}\endgroup

\end{document}